\documentclass[iop]{emulateapj}

\usepackage{natbib}
\usepackage{amsmath}
\usepackage{hyperref}

\usepackage{soul}
\usepackage[usenames]{color}
\definecolor{Blue}{rgb}{0,0.08,0.95}
\definecolor{Red}{rgb}{0.95,0.08,0.0}
\definecolor{Faint}{rgb}{0.2,0.95,0.95}

\newcommand{\omitted}[1]{}

\begin{document}

\title{Numerical simulations of active region scale flux emergence: From 
  spot formation to decay}
\shorttitle{Active region scale flux emergence simulations}

\author{M. Rempel\altaffilmark{1} and  M.C.M. Cheung\altaffilmark{2}}

\shortauthors{Rempel \& Cheung}
 
\altaffiltext{1}{High Altitude Observatory,
    NCAR, P.O. Box 3000, Boulder, Colorado 80307, USA}
\altaffiltext{2}{Lockheed Martin Solar and Astrophysics Laboratory,
  3251 Hanover St, Palo Alto, CA 94304, USA}

\email{rempel@ucar.edu}

\begin{abstract}
We present numerical simulations of active region scale flux emergence covering a time span of up to 6 days. Flux emergence is driven by a bottom boundary condition that advects a semi-torus of magnetic field with $1.7 \times 10^{22}$~Mx flux into the computational domain. The simulations show that, even in the absense of twist, the magnetic flux is able the rise through the upper $15.5$ Mm of the convection zone and emerge into the photosphere to form spots. We find that spot formation is sensitive to the persistence of upflows at the bottom boundary footpoints, i.e. a continuing upflow would prevent spot formation. In addition, the presence of a torus-aligned flow (such flow into the retrograde direction is expected from angular momentum conservation during the rise of flux ropes through the convection zone) leads to a significant asymmetry between the pair of spots, with the spot corresponding to the leading spot on the Sun being more axisymmetric and coherent, but also forming with a delay relative to the following spot. The spot formation phase transitions directly into a decay phase. Subsurface flows fragment the magnetic field and lead to intrusions of almost field free plasma underneath the photosphere. When such intrusions reach photospheric layers, the spot fragments. The time scale for spot decay is comparable to the longest convective time scales present in the simulation domain. We find that the dispersal of flux from a simulated spot in the first two days of the decay phase is consistent with self-similar decay by turbulent diffusion.
\end{abstract}


\received{}
\accepted{}

\maketitle
\section{Introduction}
Active Regions (ARs) are prominent centerpieces of solar magnetic activity. In themselves, ARs exhibit interesting manifestations of the complex interplay between magnetohydrodynamics  (MHD) and radiative transfer. As demonstrated by numerical simulations in recent years, this interplay is responsible for a wide range of phenomenological features associated with ARs, including sunspot umbrae~\citep{Schuessler:UmbralConvection,Rempel:SunspotStructure,Rempel:SunspotSubsurfaceStructure} and penumbrae~\citep{Heinemann:PenumbraFineStructure,Rempel:SunspotStructure,Rempel:PenumbralStructure,Rempel:PenumbralFineStructure,Kitiashvili:SeaSerpentPenumbra,Rempel:LivingReviews,Rempel:Robustness}, light bridges~\citep{Cheung:ARFormation}, pores~\citep{Cameron:Pores,Kitiashvili:Spontaneous,SteinNordlund:ARFormation}, plages~\citet{Voegler:MURaM} as well as anomalous features in solar granulation associated with emerging flux~\citep{Cheung:FluxEmergenceInGranularConvection,Cheung:SolarSurfaceEmergingFluxRegions,MartinezSykora:TwistedFluxEmergence,TortosaAndreu:FluxEmergence,Fang:DynamicCoupling}.

Even before 3D numerical simulations addressing AR formation were feasible, it was generally accepted that ARs form when somewhat coherent, buoyant bundles of these toroidal fields rise toward the surface~\citep[see reviews by][and references therein]{Moreno-Insertis:Review,Fisher:SolarDynamoAndEmergingFlux,Fan:LivingReview,Fan:LivingReview2009,Archontis:ReviewArticle}. The physical mechanisms that influence AR tilts and asymmetries have been addressed in idealized models that use the thin flux tube or analestic approximations. The assumptions employed in these models are not appropriate for the near-surface layers of the convection zone ($z \lesssim -20$ Mm) and 3D, fully-compressible MHD simulations are required to investigate how ARs form at whitthe surface.

Radiative MHD simulations of AR formation following flux emergence by~\citet{Cheung:ARFormation} and~\citet{SteinNordlund:ARFormation} have partially bridged this gap in our models of the life cycle of ARs but outstanding questions remain. In both models, the magnetic field is given a chosen strength and orientation at the bottom boundary and the ensuing evolution is dependent on the choice. In the model of~\citet{SteinNordlund:ARFormation}, horizontal magnetic field of constant strength and orientation is fed into the domain through the bottom boundary in upflow regions. In the case of~\citet{Cheung:ARFormation}, the magnetic configuration was chosen to be a semi-torus to mimic the rise of a somewhat coherent flux loop from layers below the computational domain. In both cases,  the origin of the magnetic fields that are fed into the computational domains is not addressed, although the existence of buoyant magnetic flux tubes coherent over length scales exceeding tens of megameters is found to be self-consistently generated in recent global dynamo simulations~\citep{Nelson:DynamoLoops}.

In the present work, we extend the approach taken in~\citet{Cheung:ARFormation} to examine how buoyant flux tubes would emerge and evolve should they be present in the convection zone. In this article, we present numerical experiments based on radiative MHD that address certain aspects of the formation and decay of ARs. In one of the numerical experiments, we follow the full life cycle of the modeled AR (i.e. from pre-emergence to decay) over the course of $6$ days. We complement this simulation with control experiments that examine the role of the bottom boundary condition we use to initiate flux emergence. The combined results allow us to examine the robustness of the AR formation process and to examine the influences that lead to AR fragmentation. In particular we aim to address the following key questions: 1. How does magnetic field decouple from the enclosed mass to allow for flux emergence and spot formation in a highly stratified medium? 2. What is the role of subsurface flows in AR evolution? 3. What are the photospheric signatures of field aligned flows that are expected to occur as consequence of angular momentum conservation? 4. Which processes govern the decay of flux concentrations after flux emergence?

The remainder of the article is structured as follows. Section~\ref{sect:setup} presents the setup of the numerical simulations. Section~\ref{sect:results} presents results and lessons learned from the numerical simulations. Finally, section~\ref{sect:discussion} discusses the implications of our findings in the context of emerging flux regions, AR formation and the solar dynamo.

\section{Numerical setup}
\label{sect:setup}
The numerical simulations presented here is computed with the MURaM radiative
MHD code \citep{Voegler:MURaM,Rempel:SunspotStructure}. The domain size is
$147.456\times 73.728\times 16.384$ Mm$^3$ at a resolution of 
$96\times 96\times 32$ km$^3$, leading to a grid size of 
$1536\times 768\times 512$ grid points. The domain is periodic in horizontal
directions, open for upflows at the top boundary and open at the bottom
boundary. The open lower boundary is implemented through a symmetric condition
on all mass flux components (i.e. values of mass flux are mirrored across the boundary), 
while the pressure is kept fixed at the boundary to maintain the
total mass in the domain (the pressure is extrapolated linearly into the boundary cells such that the pressure at the interface
between the first boundary and first domain cells remains constant). 
The specific entropy is symmetric in downflow regions. In upflow regions, the entropy is specified such that the resulting radiative losses in the photosphere
lead to a solar-like energy flux of about $6.3\cdot 10^{10}\,\mbox{erg}\,\mbox{cm}^{-2}\mbox{s}^{-1}$ through the domain under quiet-Sun conditions. 
Magnetic field is vertical ($B_x$, $B_y$ antisymmetric, $B_z$ symmetric) at the bottom and matched to a potential field extrapolation at the top.

We start our simulations from a snapshot with thermally relaxed, non-magnetic convection.
Magnetic flux emergence is initiated by kinematically advecting a 
semi-torus of magnetic field across the bottom boundary.  To this end the above mentioned boundary condition
is overwritten in the region with flux emergence by specifying a velocity and magnetic field in the boundary cells following the approach
detailed in \citet{Cheung:ARFormation}. Within the flux emergence region we extrapolate the pressure hydrostatically into the 
boundary layers, while we keep the fixed pressure boundary condition outside. This allows for a pressure adjustment within the flux emergence region in response to the imposed inflow at the boundary. The entropy is set to be equal to the value we impose in upflows outside the flux emergence region. Our setup differs from previous work in the following aspects:

With the domain size we scaled up the proportions of the torus shaped flux loop compared to our previous work. The major radius of the torus is $R=24$ Mm, the minor radius is $a=7.64$ Mm. The cross sectional profile of the torus is a Gaussian, which is cut off at an amplitude of $0.135$ (radius of $r=\sqrt{2}a=10.8$ Mm) in order to have a well defined outer boundary. With a  field strength of $10.6$ kG at the axis of the torus, the total flux contained in the emerging flux loop is $1.7\cdot 10^{22}$ Mx. The torus is initially located such that its center is $R+\sqrt{2}a=34.8$ Mm below the bottom boundary. Starting at $t=0$, the torus is kinematically advected across the bottom boundary of the domain with a vertical velocity
of $500\,\mbox{ms}^{-1}$. The total time required for the semi-torus to be advected across the bottom boundary is $19.33$ hours. After the semi-torus has been advected through the bottom boundary ($t=19.33$ hours), we smoothly transition over a time  interval of $2.77$ hours to the open boundary as described above (during the transition phase the boundary condition is written as a
linear combination of flux emergence and open boundary condition with a linear transition in time). Starting from $t=22.1$ hours the
boundary condition is again open everywhere at the bottom boundary. The overall time scale for flux emergence in our setup is not very different from a convective time scale. Based on the average convective upflow velocity it takes
about $15$~hours for upflowing material entering at the bottom boundary to reach the photosphere.

\begin{figure*}
  \centering 
  \resizebox{0.95\hsize}{!}{\includegraphics{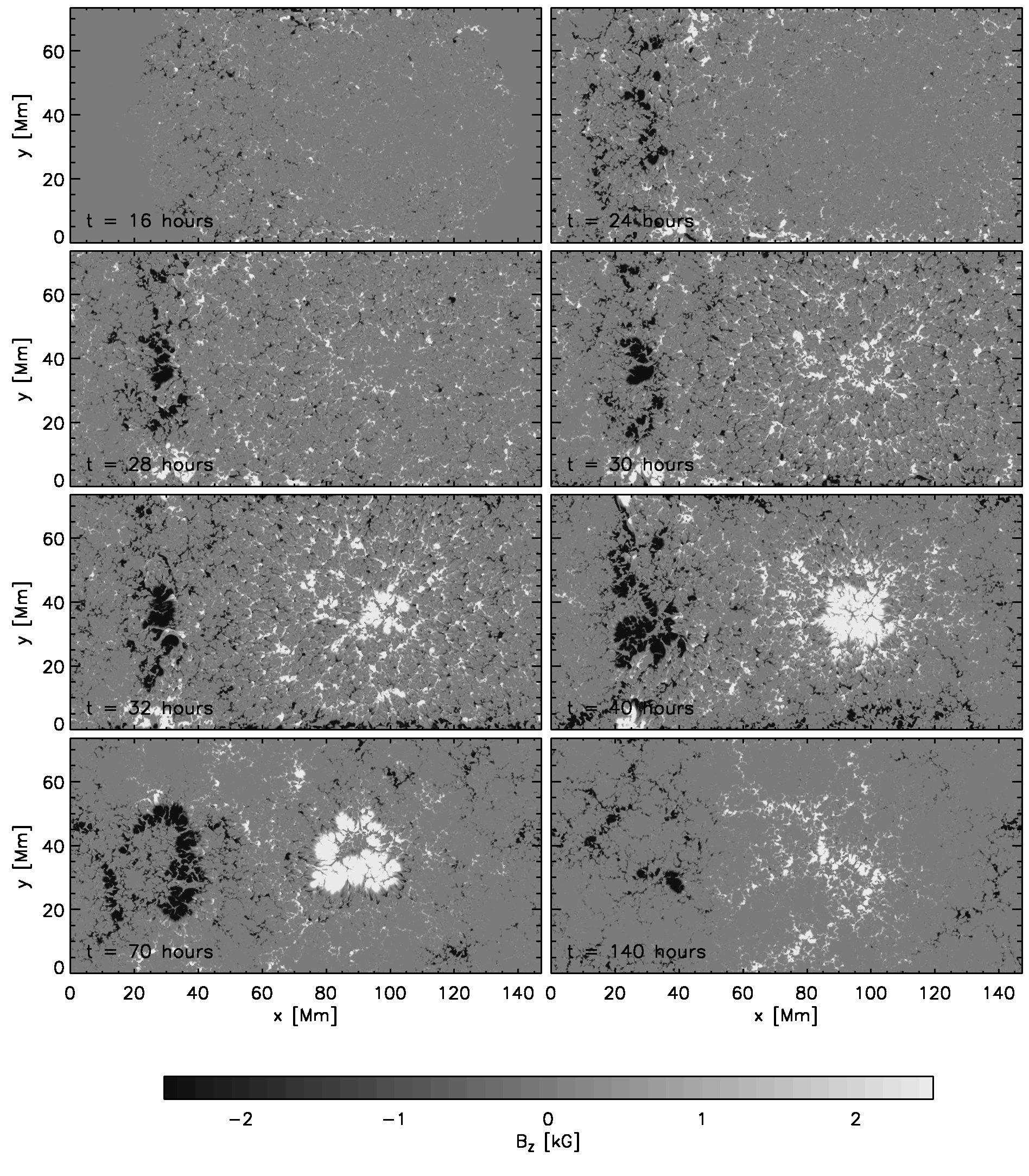}}
  \caption{Time evolution of $B_z$ at the $\tau=1$ level in the
    photosphere. We present snapshots at $t=16, 24, 28, 30, 32, 40, 70$
    and $140$ hours. For comparison the flux emergence process at
    the bottom boundary is finished at $t=19.33$ hours, the transition
    to an open boundary condition at $t=22.1$ hours. An animation 
    combining Figures 1 and 3 is provided in the online material.
  }
  \label{fig:f1}
\end{figure*}

\begin{figure}
  \centering 
  \resizebox{0.95\hsize}{!}{\includegraphics{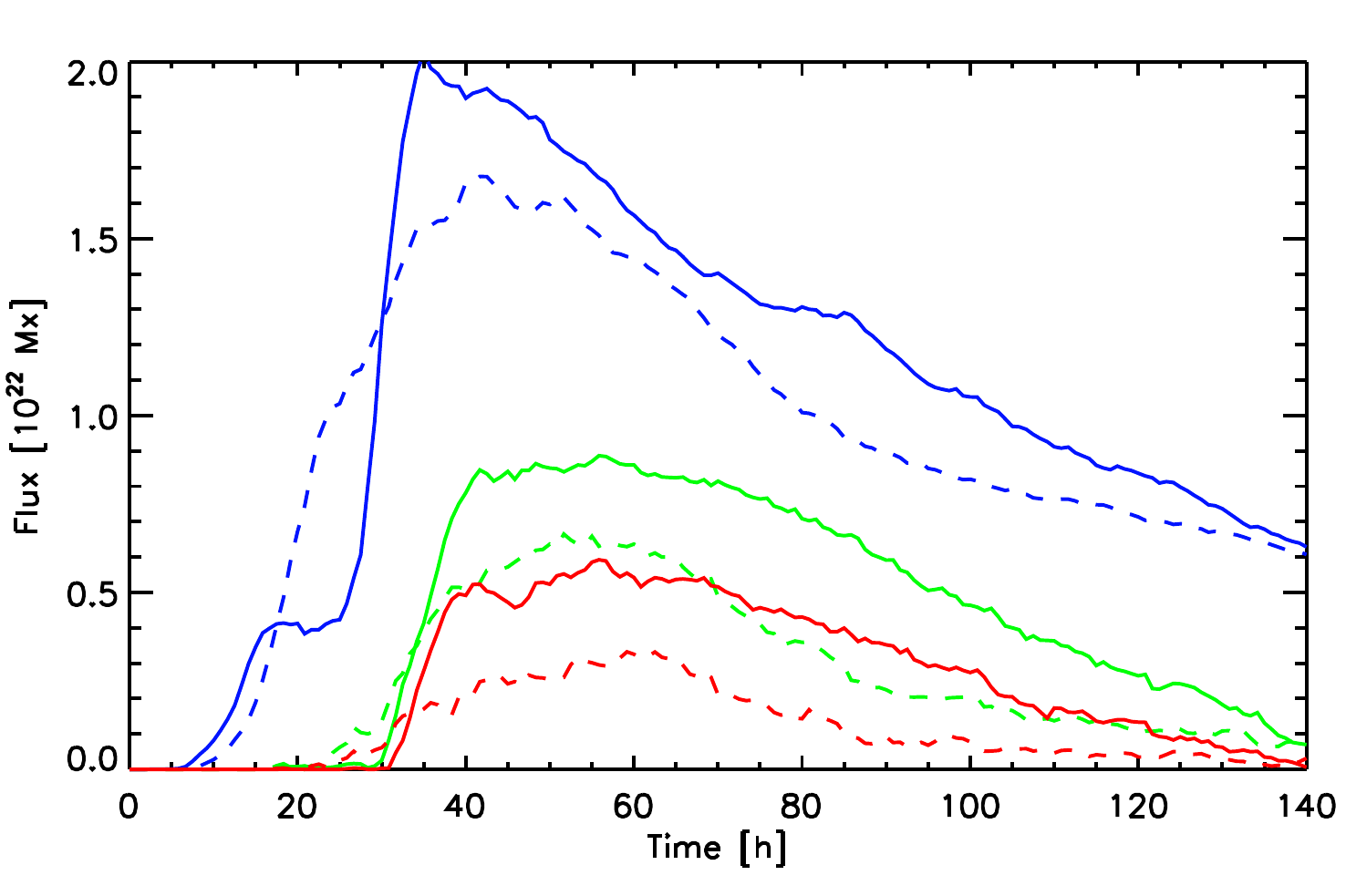}}
  \caption{Time evolution of unsigned flux at $\tau=1$ (blue) as well
    as signed flux in regions with $\overline{I}<0.8 I_{\odot}$ (green) and
    $\overline{I}<0.5 I_{\odot}$ (red). Solid lines correspond to the right spot,
    dashed lines correspond to the left spot. 
  }
  \label{fig:f2}
\end{figure}

In addition to the setup described above we added in the simulation described 
here a field aligned flow of $500\,\mbox{ms}^{-1}$ with a Gaussian cross 
section. This choice of an azimuthal flow within the torus is motivated by results from simulations of global toroidal flux tubes rising through the convection zone~\citep{Fan:ThinFluxTubeI,Fan:ThinFluxTubeII,Moreno-Insertis:pf-assymetries,Caligari:EmergingTubesPartI,Caligari:EmergingTubesPartII,Fan:FatFluxTubes,Weber:RiseOfARFluxTubes,Fan:LessFatFluxTubes}. For instance, 3D anelastic simulations of the rise of global toroidal flux tubes through a turbulent rotating solar convection zone yields retrograde flows at the apexes exceeding $300\,\mbox{ms}^{-1}$~\citep{Fan:LessFatFluxTubes}. This arises from angular momentum conservation as flux is transported from the base of the convection zone toward the surface (increasing its distance from the axis of rotation). In our setup
this flow is directed in the negative $x$-direction, implying that our
right spot corresponds to the leading spot on the Sun. Our setup implies 
that toward the end of out flux emergence process $t=19.33$ hours the right
leg of the torus has an upflow of $1\,\mbox{km s}^{-1}$ (in the center), while 
the upflow mostly diminished at the left leg. We have chosen here a peak flow speed of $500\,\mbox{ms}^{-1}$ to be comparable to the speed of flux emergence. This choice was made to maximize the potential effect, but is not too different from results found in the above mentioned flux emergence simulations. Since the thin flux tube approximation assumes no variation of quantities over the cross section of the tube, the quantity to be compared is the flow speed
averaged over the cross section. With the Gaussian profile truncated at a value of $0.135$, the mean flow speed is $220\,\mbox{m s}^{-1}$. 

Discussion of the simulation results will focus on the simulation run with the setup described above. However, it will be complemented with results from other control experiments. These control experiments allow us to examine the robustness of the spot formation mechanism. They do not have the field-aligned flow and have different initial field strengths and bottom boundary conditions. We describe these simulations in more detail in Section \ref{sect:controlexp}.

\subsection{Scope of the simulations presented here}
\label{sect:setup:scope}
Active region formation and decay is a multi length- and time-scale
problem. It is currently not feasible to address all involved processes in one
numerical simulation. With the setup presented here we focus on large-scale
and long-term evolution aspects at the expense that we cannot cover for example
details related sunspot fine-structure, such as penumbra formation and 
evolution. It was shown by \citet{Rempel:Robustness} that penumbral fine-structure
requires substantially higher resolution than we use here and in addition
also a sufficiently inclined magnetic field at the top boundary. Our
potential field extrapolation in combination with horizontal periodicity and
a top boundary about $700$ km above the photosphere is insufficient in that regard. 
Addressing penumbra formation requires likely also a more sophisticated treatment of 
the layers overlying the photosphere, which is beyond the scope of this investigation. 

In our setup the flux emergence process is driven through the bottom boundary and as a consequence
several aspects of the simulation (which we describe in further detail later) are dependent
on that boundary condition. In that regard the work presented here should be considered as
``numerical experiments'' to explore the role of flow fields in $10-15$~Mm depth. The aim of this 
paper is to present a prototype simulation that highlights how a combination of upflows transporting 
flux toward the photosphere and flows along that flux system (as suggested by several rising flux tube
simulations) can influence the photospheric field evolution. The values for these two
flow components chosen here (together with the initial field strength) are an ``educated guess'' based on 
previous modeling results, but they are by no means intended to represent the conditions of a typical
active region on the sun. Constraining the latter is the ultimate goal and simulations like the one presented
here can provide some guidance in that regard as a forward modeling tool. We have performed several simulations 
in addition to what we present here (with different assumptions about field strength and flow amplitudes) and a 
detailed comparison with observations is work in progress.

Since the goal of this paper is not the most realistic representation of an active region we have also
chosen for simplicity to consider only untwisted magnetic field. We discuss in Section \ref{sect:penumbra_twist}
potential consequences of not including twist and a penumbra in our simulation setup.

\section{Results}
\label{sect:results}
\subsection{Flux emergence and spot formation}

\begin{figure*}
  \centering 
  \resizebox{0.95\hsize}{!}{\includegraphics{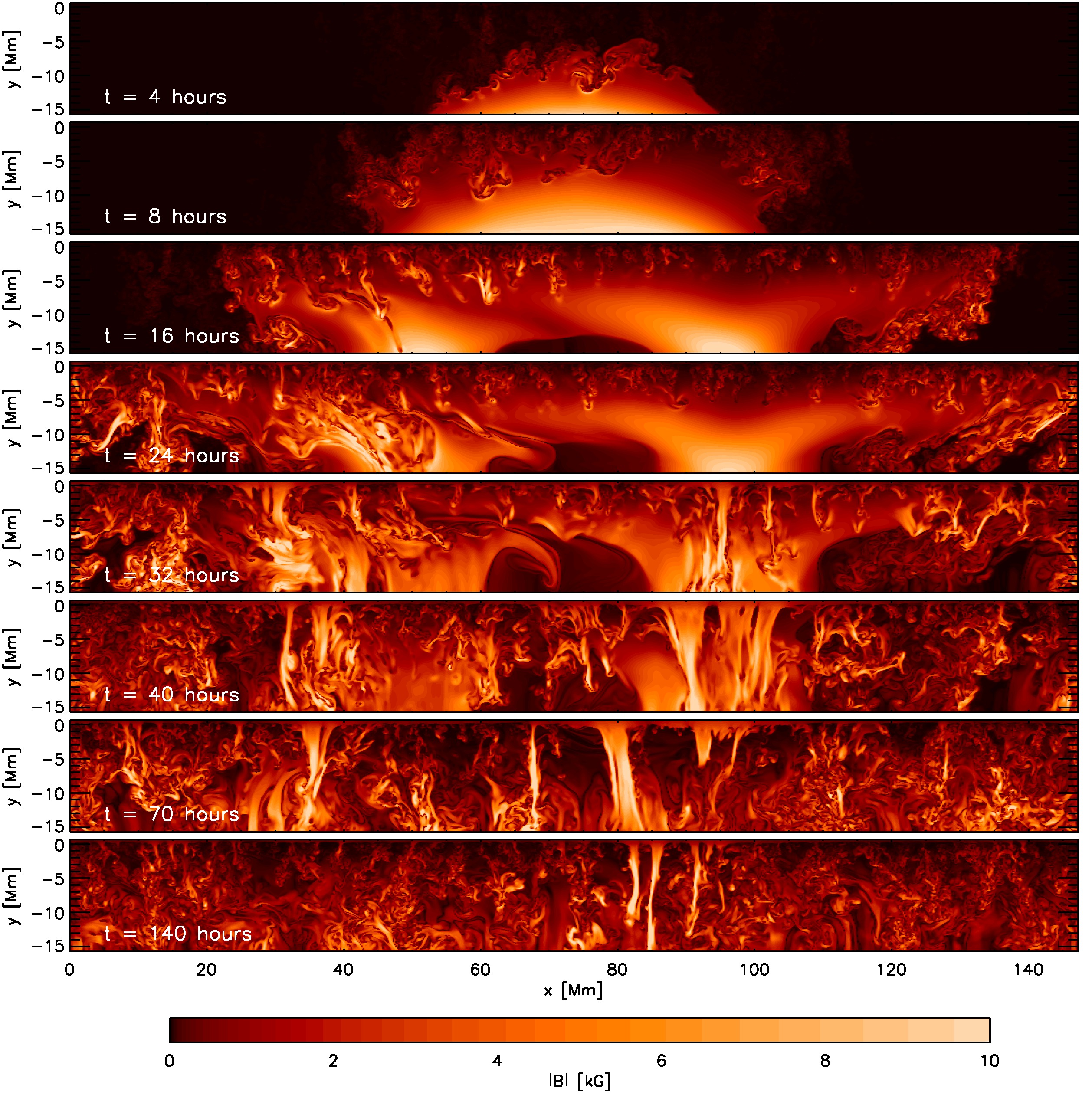}}
  \caption{Time evolution of $\vert B\vert$ on a vertical cut through the
    center of the domain along the x-axis. The first two snapshots show
    the subsurface field evolution prior the appearance of flux in
    the photosphere, the remaining six snapshots correspond to the
    photospheric magnetograms shown in Fig. \ref{fig:f1} (except for those at 
    $t=28$ and $30$~hours). An animation 
    combining Figures 1 and 3 is provided in the online material.
  }
  \label{fig:f3}
\end{figure*}

\begin{figure*}
  \centering
  \includegraphics[width=0.6\textwidth]{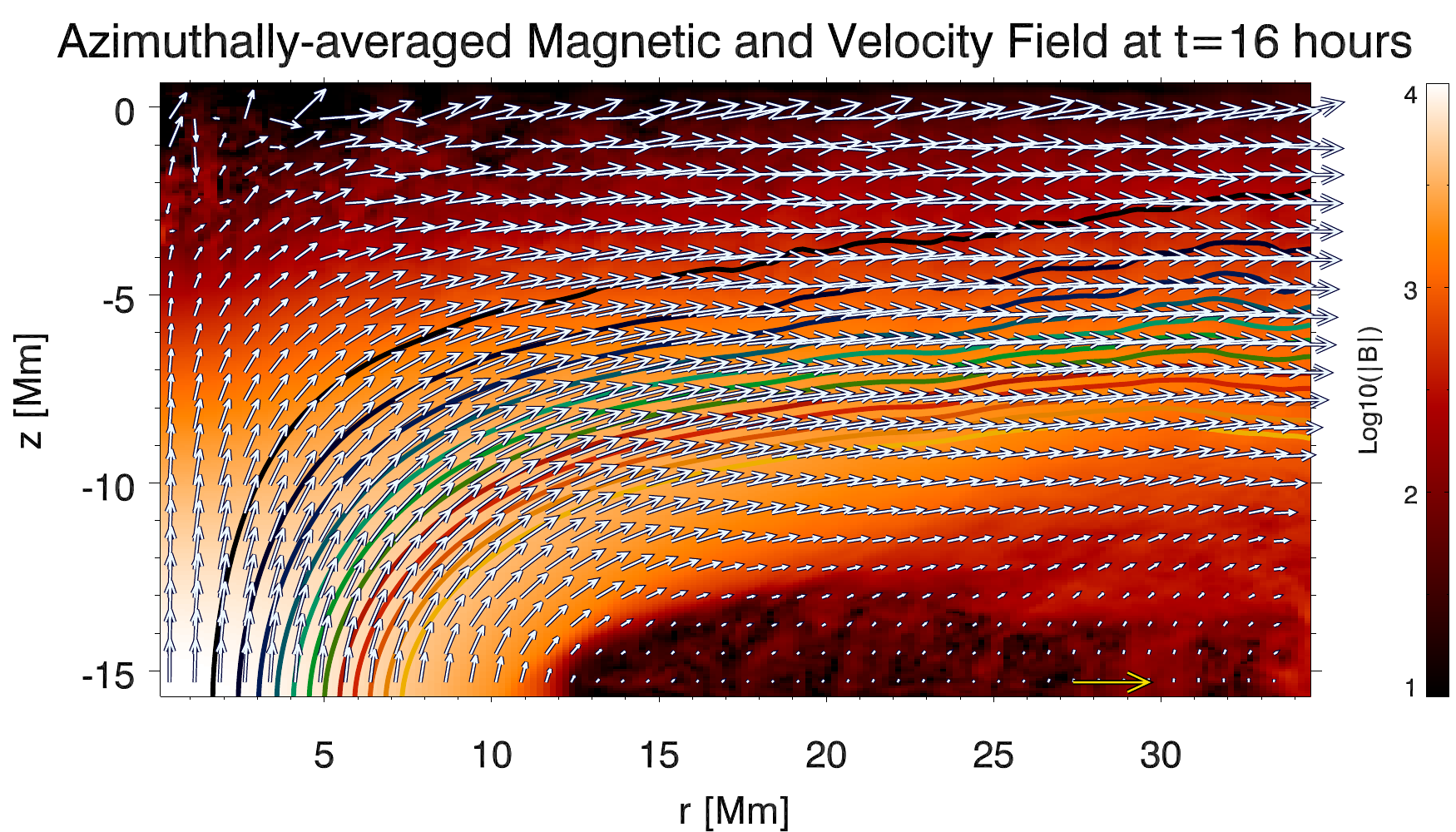}
  \caption{Subsurface structure of the developing right spot at $t=16$ hours. The azimuthally-averaged magnetic field is $\overline{\mathbf{B}}$ shown in orange ($\|\overline{\mathbf{B}} \|$ in logarithmic scale from $10$ G to $10$ kG). The contour lines show flux surfaces enclosing flux content ranging from $\Phi=10^{21}$ to $\Phi=1.2\times 10^{22}$ Mx in intervals of $\Delta \Phi = 10^{21}$ Mx. Overlaid vectors display components of the azimuthally-averaged velocity field $\overline{\mathbf{v}}$ in the $r$-$z$ plane. The yellow arrow at the bottom of the plot corresponds to 2 km/s. The average flow field is roughly aligned with the average magnetic field, which loads the subsurface layers with buoyant, predominantly horizontal magnetic flux ready to emerge.}
  \label{fig:f4}
\end{figure*}

\begin{figure*}
  \centering 
  \resizebox{0.95\hsize}{!}{\includegraphics{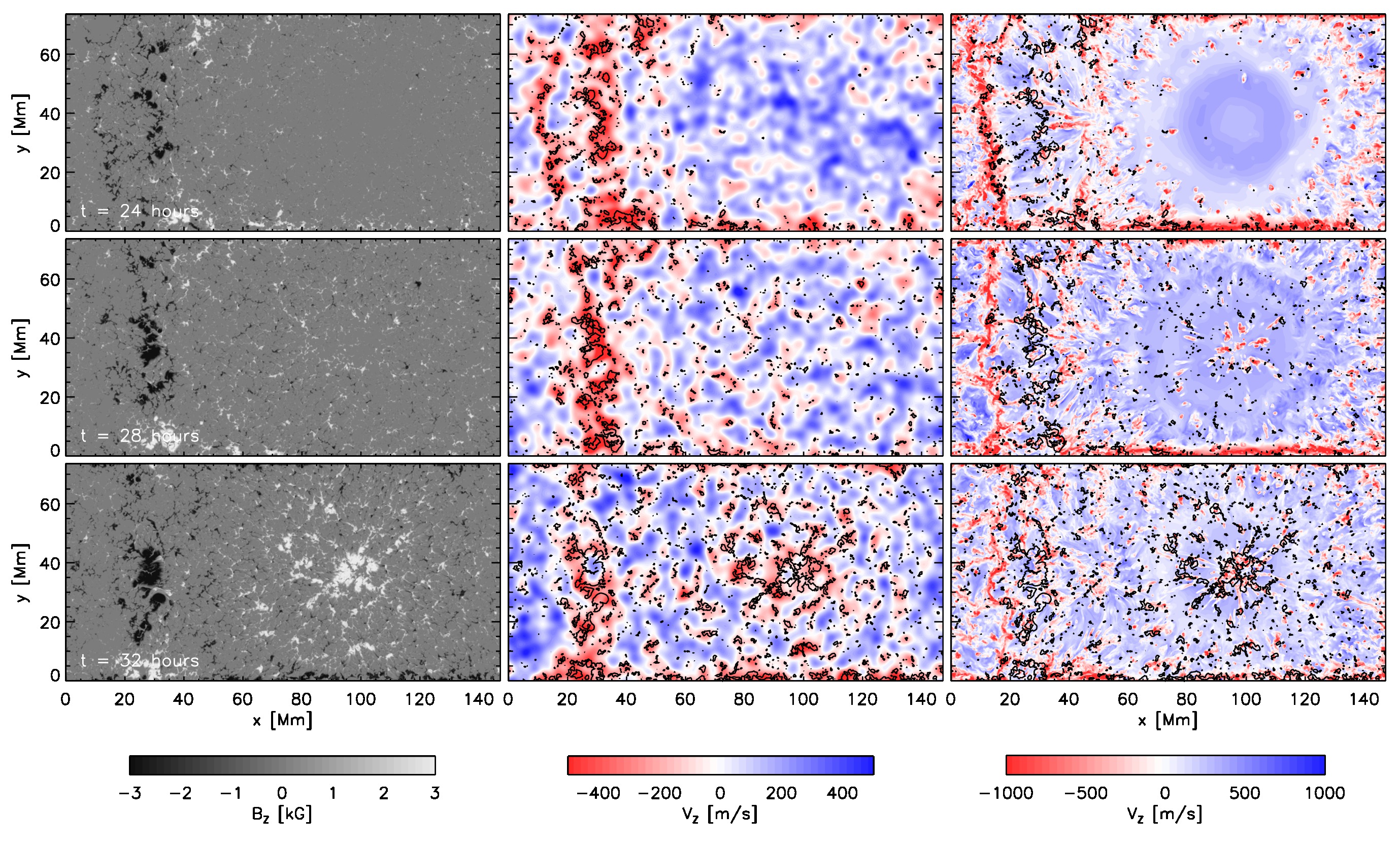}}
  \caption{Magnetograms at $\tau=1$ (left),  smoothed vertical velocity at 
      $\tau=1$ (middle), and vertical flow velocity in
    $8$~Mm depth (right) for three different times during the spot formation 
    phase ($t=24, 28$ and $32$~hours). In the velocity plots black contours 
    indicate vertical magnetic field with more than $1.5$~kG in the photosphere. 
    Positive velocity values (blue colors) indicate upflows.
  }
  \label{fig:f5}
\end{figure*}

\begin{figure*}
  \centering 
  \resizebox{0.8\hsize}{!}{\includegraphics{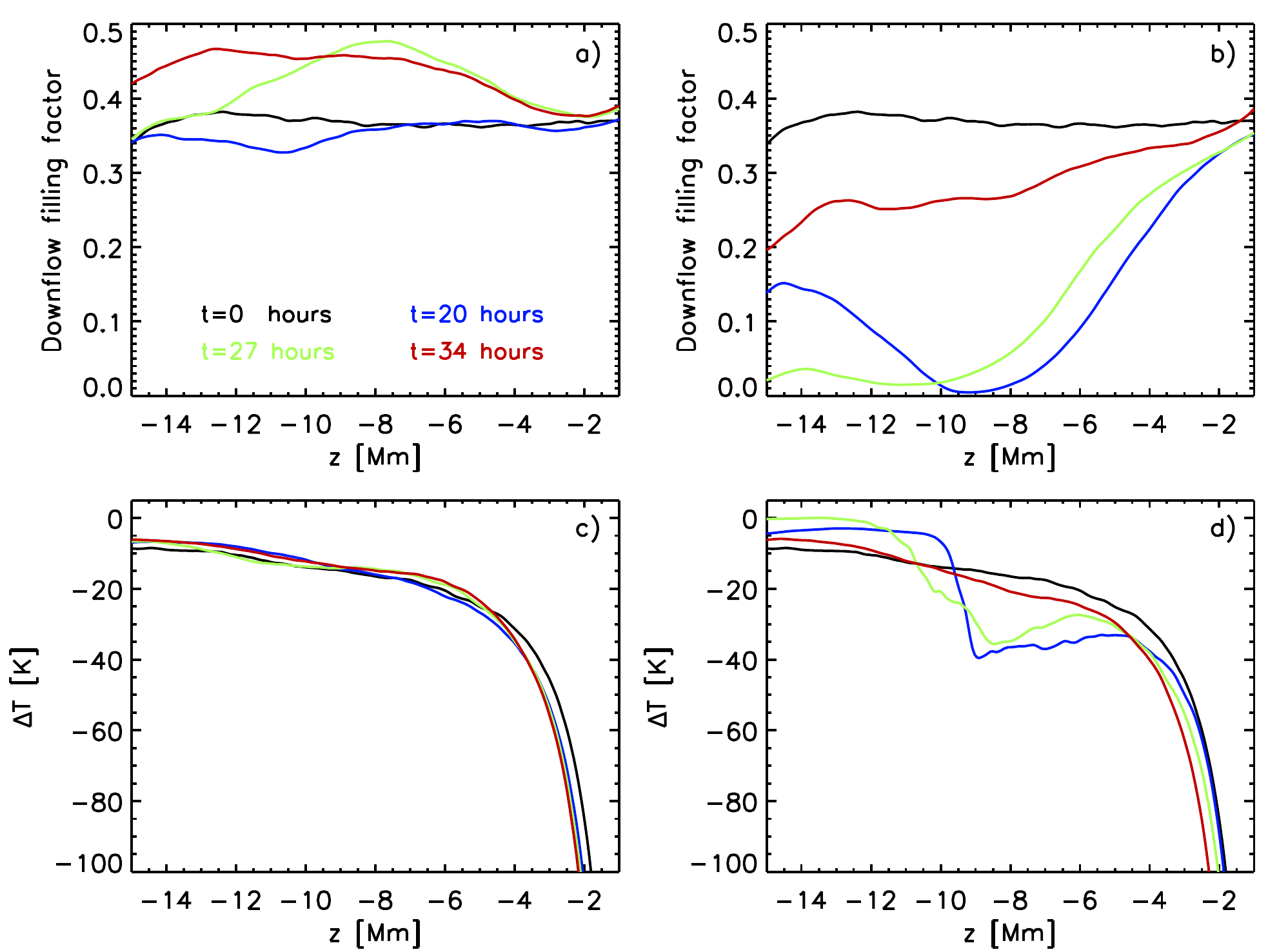}}
  \caption{Panels a) and b): Vertical profiles of downflow filling factors.
    Panels c) and d): Temperature difference $\Delta T=T_{down}-T_{up}$
    between downflow and upflow regions. Except for the snapshot at $t=0$~hours
    (black line), which is based on an average over the whole domain, the 
    quantities are computed inside a cylinder with $25$~Mm radius centered on 
    the left spot (a, c) and right spot (b, d). In addition we averaged
    2~hours in time.
  }
  \label{fig:f6}
\end{figure*}

\begin{figure*}
  \centering 
  \resizebox{0.95\hsize}{!}{\includegraphics{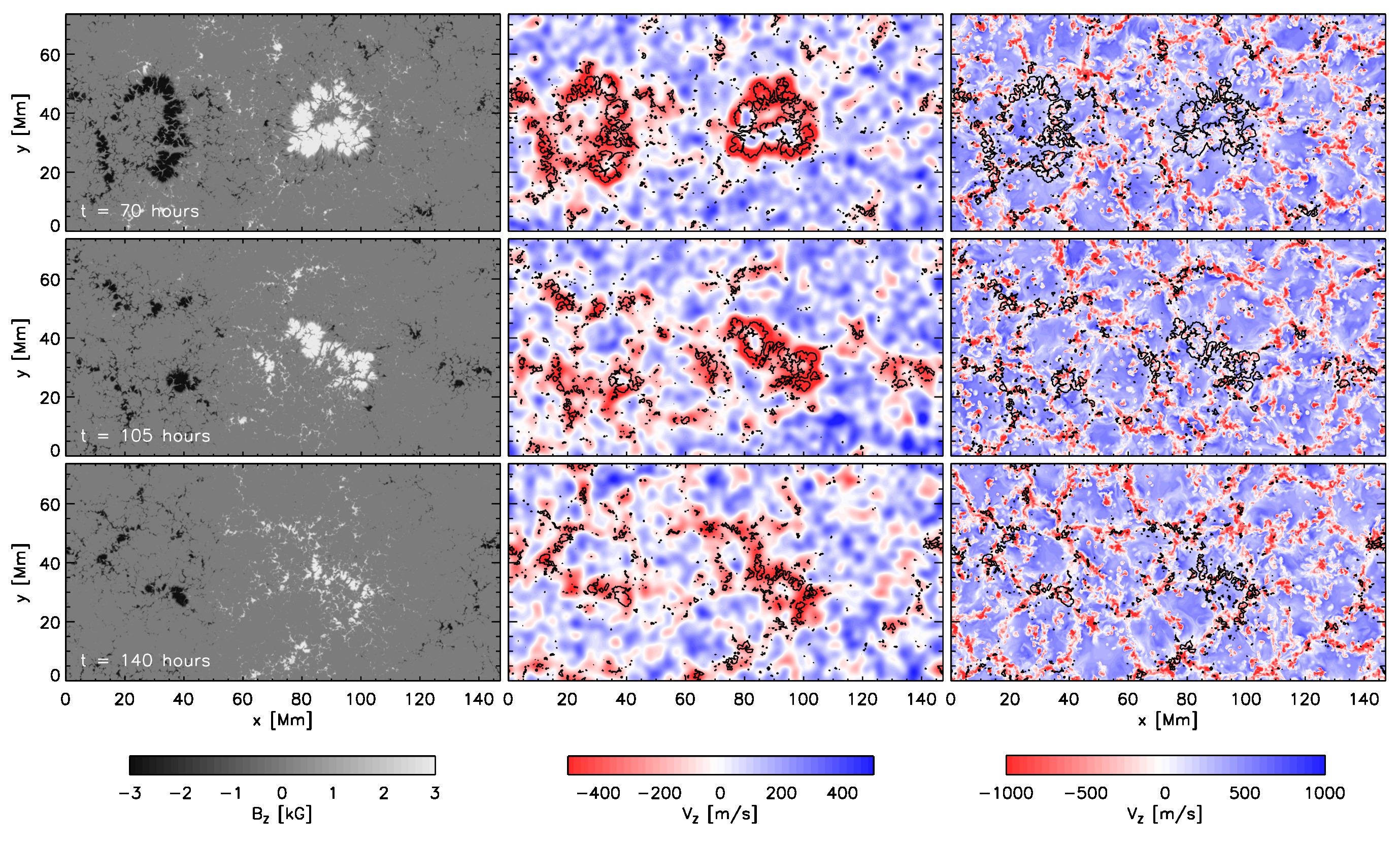}}
  \caption{Magnetograms at $\tau=1$ (left), smoothed vertical velocity at 
      $\tau=1$ (middle) and vertical flow velocity in
    $12$~Mm depth (right) for three different times during the active region 
    decay phase. In the velocity plots black contours indicate vertical 
    magnetic field
    with more than $1.5$~kG in the photosphere. The shape of the decaying 
    spots and distribution of strong flux concentrations in the
    plage region is determined to a large degree by the convection patterns
    near the bottom of the domain. 
    Positive velocity values (blue colors) indicate upflows.
  }
  \label{fig:f7}
\end{figure*}

\begin{figure*}
  \centering 
  \resizebox{0.95\hsize}{!}{\includegraphics{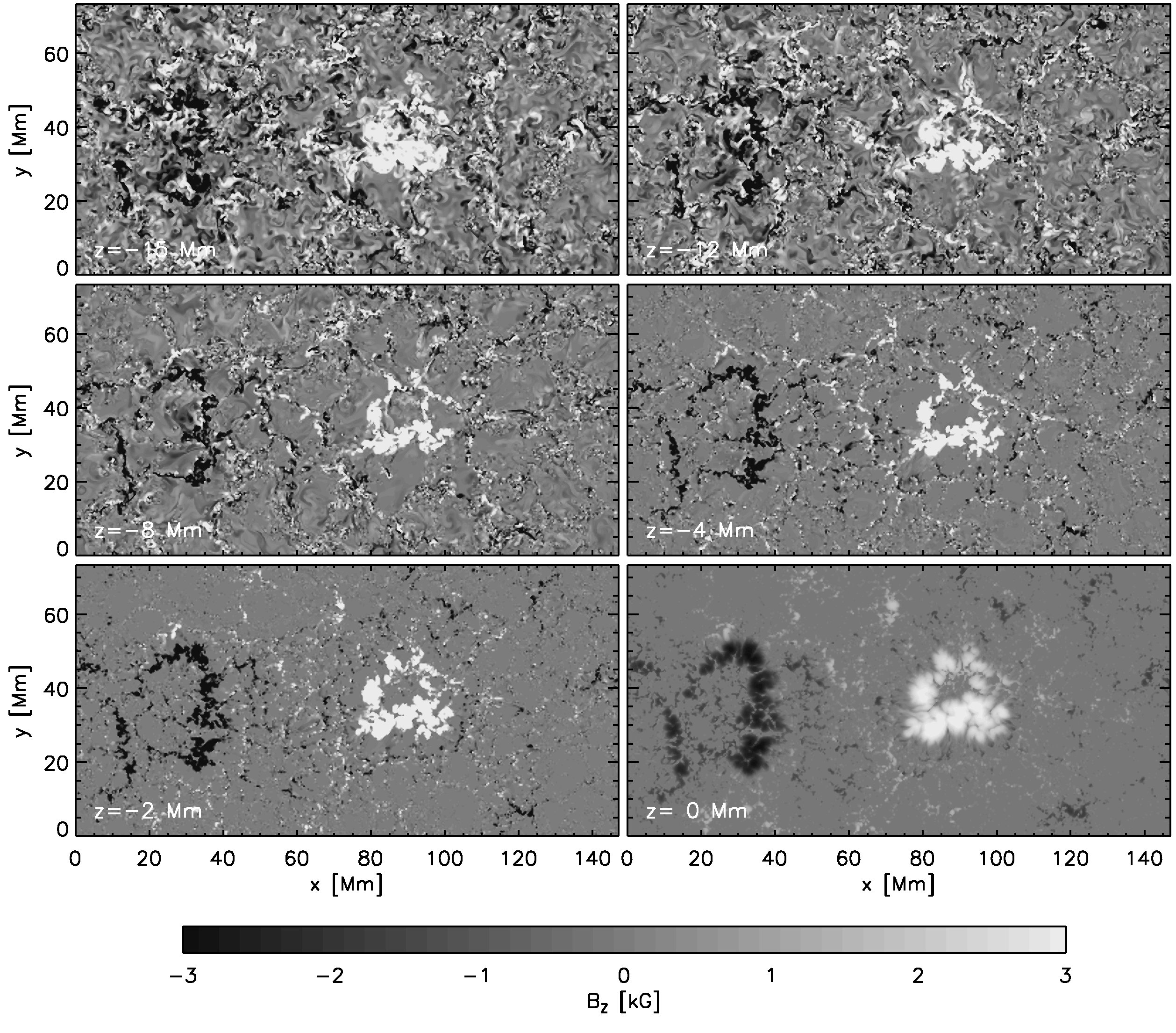}}
  \caption{Vertical magnetic field at the six depth levels
    $z=-15$, $-12$, $-8$, $-4$, $-2$, and $0$~Mm. The field strength
    is saturated at $3$~kG. Strong fields shows a large degree of coherence
    throughout the domain. Shown is data for the $t=70$~hours snapshot.
    An animation is provided in the online material.
  }
  \label{fig:f8}
\end{figure*}

We first present results from the simulation with the field-aligned flow along the torus.
Figure \ref{fig:f1} presents 8 snapshots displaying the evolution of the
photospheric magnetic flux through the flux emergence and subsequent
decay process. Since it is difficult to capture the time evolution
in this simulation in $8$ snapshots we provide a movie combining Figure \ref{fig:f1}
and \ref{fig:f3} in the online material.
The snapshot at $t=16$ hours shows the structure of the first 
flux appearing in the photosphere. Despite the large degree of coherence we 
impose at the bottom boundary, the flux is heavily distorted on its way through
the convection zone and organized on a  granular scale with mostly
mixed polarity when appearing first in the photosphere. It was shown by
\citet{Cheung:ARFormation} that the emerging field weakens as it rises as
$B\sim\sqrt{\varrho}$, which translates in our setup with a density contrast
of about $10^4$ to a reduction in field strength of about a factor of $100$, 
i.e. the resulting average field in the photosphere is less than $100$~G.
Asymmetry with respect to the $x$-direction becomes visible early on as a 
consequence of the field aligned flow (right to left), which leads to an earlier concentration of flux in the left half of the domain. At $t=24$ hours the field accumulation in the left half of the domain becomes strong enough to form
the first $\sim 3$ kG field concentrations, while the right half of the domain remains mostly field free at the photosphere. 
This changes starting from about $t=28$~hours through the emergence of mixed
polarity field followed by a separation of polarities and formation of a more
coherent spot on the right from $t=32$ to $t=40$ hours.
The remaining two snapshots at $t=70$ and $t=140$ hours show the decay phase
of the active region. The decay is primarily due to fragmentation of flux,
which is caused by subsurface flows (see Section \ref{sect:subsurf}). At 
later stages flux becomes organized in cells that are in size comparable to 
supergranular scales. 

Figure \ref{fig:f2} shows the evolution of unsigned flux in the photosphere (blue) as 
well as signed flux in regions with $\overline{I}<0.8 I_{\odot}$ (green) and
$\overline{I}<0.5 I_{\odot}$ (red). Here $\overline{I}$ refers to an intensity
smoothed by a Gaussian with FWHM of $1$~Mm to focus the latter two measures on larger scale
coherent flux concentrations. We separated the contributions from the left and right half of
the domain at $x=60$~Mm (this is not the center of the domain since the field aligned flow in
our setup leads to a transport of flux in the negative $x$-direction). Solid and dashed lines correspond to the right and left halves, respectively.
The emergence of mixed polarity flux is evident from this
figure, since in both halves of the domain the unsigned flux (blue) rises before the signed flux
in regions with reduced intensity (green and red). The unsigned flux reaches a value of about twice
the flux found later in the spots before the spot formation starts. This indicates that spot
formation is mostly a reorganization of flux already present in the photosphere through a separation
of polarities. The right spot is forming very rapid within about $10$~hours, while
the left spot shows a more gradual increase in flux content over a time frame of $20-30$~hours. The
formation of the left spot starts before the right spot, although the flux content present in areas 
with ($\overline{I}<0.8 I_{\odot}$) does not exceed $10^{21}$~Mx at the time the right spot starts
forming. In terms of the total flux content (measured by the unsigned flux) we find an immediate transition from formation into decay. The signed flux of the right spot shows a more or less stationary phase lasting about $30$~hours before the decay phase starts. 

Figure \ref{fig:f3} shows the evolution of the magnetic field strength on
a vertical cut through the center of the domain at $y=36.864$~Mm. The
first two snapshots ($t=4$ and $t=8$ hours) show the evolution prior to
the emergence of flux in the photosphere. The remaining six snapshots
correspond to the times shown in Figure \ref{fig:f1} (except for $t=28$ and $32$ hours).
The initial two snapshots exhibit only slight signs of asymmetry. The lower 
two thirds of the domain show a very coherent flux rope, while the upper third
shows the influence from vigorous convection distorting the field. Recent numerical studies of flux emergence in granular convection have reported similar behavior~\citep{Cheung:FluxEmergenceInGranularConvection,Cheung:SolarSurfaceEmergingFluxRegions,MartinezSykora:TwistedFluxEmergence,Fang:FluxEmergence}.

A clear asymmetry between the right and left spots develops in the snapshots at $t=16$ and $t=24$ hours. While the right spot shows a very symmetric funnel-shaped field structure in response to the stronger inflow at the right foot-point, the field in the left half of the domain is less coherent. The funnel-shaped field structure on the right bares resemblance to the subsurface field structure in our previous simulation of AR formation~\citep[see Fig. 2 of][]{Cheung:ARFormation}. It is a consequence of the mass inflow at the bottom boundary condition and the strong density stratification of the domain.  The mass density near the top of the domain is too small to carry the imposed mass flux, which, as a consequence, has to turn over at several Mm depth. This is a general property of stratified convection. It was found by \citet{Trampedach:2011:MassMixing} that the mass flux mixing length in stratified convection experiments is about $1.8$ pressure scale heights. For the domain depth considered less than $0.1\%$ of the mass flux present near the bottom of the domain reaches the photosphere.

Due to the frozen-in condition magnetic field lines must, on average, follow the streamlines of the flow, which leads to a buildup of a layer of organized horizontal field at several Mm depth. This is illustrated in Fig.~\ref{fig:f4}, which shows the azimuthally-averaged structure of the magnetic and velocity fields ($\overline{\mathbf{B}}$ and $\overline{\mathbf{v}}$, respectively) of the right spot at $t=16$ hours. The presence of sustained outflows from the center of the developing spot, driven by mass injection through the footpoint of the torus at the bottom boundary, has the effect of preventing magnetic flux from reaching the photosphere. So while subsurface upflows are necessary to load the subsurface layers with magnetic flux, the persistence of such flows is not amenable to the formation of the spots.

During this phase we find horizontally diverging flows reaching an amplitude of up to $2\,\mbox{km s}^{-1}$. This flow amplitude is 
comparable to surface Doppler measurements during the early stages of active region formation 
\citep[e.g.][]{Toriumi:2012:HDF,Khlystova:2013:FemFlows}, although these observed flows are at least partially related to the separation of polarities and not
limited to divergent flows around one of the polarities as seen in our setup. The latter is a consequence of setting the field aligned flow component in our setup equal
to the upflow component.
On the other hand there is no helioseismic evidence for divergent flows
during the 24 hours preceding flux emergence \citep{Birch:2013:PreFEM}. Our current simulations do not capture a time period that is comparable
to their study, which is 24 hours prior to $10\%$ of the peak flux emergence rate in the photosphere.

The situation for the right spot changes after $t=32$ hours following the decay of the mass inflow through the bottom boundary. The predominantly horizontal magnetic layer with field strengths around $4-6$~kG is magnetically and thermally buoyant and starts rising toward the surface. It is only after flux has emerged into the photosphere whence the spot formation process may commence. The process by which dispersed flux organizes into coherent spots will be discussed in the following paragraphs. The time delay between the first signs of umbral formation between the right and the left spots can also be explained in terms of horizontal outflows. Whereas the right polarity region begins to develop pores at $t\sim 32$ hours, its left counterpart had already done so 8 hours prior (compare field distribution of the two polarities at $t=24$ and $t=32$ hours in Fig.~\ref{fig:f1}). This is due to the superposition of the toroidal `retrograde' flow of $500\,\mbox{ms}^{-1}$ Over the imposed vertical upflow of $500\,\mbox{ms}^{-1}$, which effectively shuts off the inflow through the footpoint on the left by $t=19.33$ hours (i.e. where the top half of the torus has been advected through the bottom boundary). We possibly overestimate the asymmetry and delay in spot formation in our setup, since the imposed 'retrograde' flow of $500\,\mbox{ms}^{-1}$ was chosen to maximize the effect. In addition the formation of the left spot is also influenced by the horizontal domain extent, which prevents further spreading of magnetic flux due to horizontal periodicity (see movie provided in the online material)

When the buoyant magnetic flux arrives at the photosphere, it emerges in the form of field lines undulated by granular convective flows. In terms of the surface magnetogram (Fig.~\ref{fig:f1}, $t=16$ to $28$ hours), these undulated field lines appears as mixed polarities structured at the granular scale. Opposite polarities systematically migrate in different directions (Fig.~\ref{fig:f1}, $t=30$ and $32$ hours), such that flux with negative polarity moves toward the edges of of the
domain, while the positive polarity forms a coherent almost axisymmetric
spot on the right (Figure \ref{fig:f1}, $t=32$ and $40$~hours). The magnetic
field of the right spot reaches field strengths in the $3-4$~kG
range in the photosphere and forms strands of strong field reaching to
the bottom boundary (Figure \ref{fig:f3}, $t=32$ and $40$~hours). At the 
same time the initially coherent foot-points of the emerged half torus start 
to disappear after $t=40$~hours. 

Figure \ref{fig:f5} present the connection between the 
photospheric vertical magnetic field (left panels) and the vertical flow velocity in 
$8$~Mm depth (right panels) during the formation phase of the spots. In addition we
present in the middle panels the smoothed vertical velocity at $\tau=1$ in order to
focus on the larger scale velocity components (we used here a convolution with as Gaussian
having a full width at half maximum of $5$~Mm). We subtracted systematic velocity offsets,
which result from the different height of the $\tau=$ surface in up- and downflow regions, 
density variations between up- and downflows and box-oscillations.
At $t=24$ hours
the subsurface velocity shows a strong asymmetry between the left and right
half of the domain. A strong almost circular upflow region is still present 
on the right side and most convective downflow lanes have been pushed away. 
Only a few isolated downflows
are present near the periphery. The left side of the domain shows several
downflow lanes with alignment in the y-direction, which are the locations
where the first flux of the left spot is amplified in the photosphere. 
At $t=28$ hours strong downflows are present in $8$~Mm depth centered in the
location above the foot-point of the right spot. These downflows precede the 
appearance of strong field in the photosphere around $t=32$ hours and are more
the cause rather than consequence of the right spot formation. In all three
snapshots the large scale photospheric vertical velocity shows downflows where magnetic 
flux accumulates.

To better 
understand the dynamical origin of these downflows we present in Figure 
\ref{fig:f6} for different time steps the downflow filling factor and
temperature difference between down- and upflows. These quantities are 
computed for a cylindrical volume with $25$~Mm radius centered on the
forming spots. The left panels correspond to the left, the right 
panels to the right spot. The black lines ($t=0$ hours) show
the convective reference prior to the flux emergence. Overall the
changes are very moderate in the case of the spot on the left. There is
a moderate increase in the overall downflow filling factor, consistent
with the fact the left spot is forming above the downflows at the periphery 
of the flux emergence region. There is no significant change
in the downflow/upflow temperature difference. The situation is very different
for the right spot. Initially we find a more or less constant 
downflow filling factor around $35-40\%$. The flux emergence causes the 
downflow filling factor to drop throughout most of the domain and downflows 
become almost completely suppressed for $z<-10$~Mm. At the same time the 
temperature contrast between down- and upflows increases sharply above 
$z=-10$~Mm: Cool downflows are continuously formed in the photosphere due to
radiative cooling, but the suppression of downflows below $z<-10$~Mm
leads to an accumulation of cool material around a depth of $8-10$~Mm.
At $t=20$ hours the temperature contrast is increased by about a factor of $3$
in $9$~Mm depth compared to the convective reference (black line).
When the inflow at the bottom diminishes, the top heavy stratification leads
to strong downflows that aid the formation of the spot on the right. At $t=27$~hours 
(begin of right spot formation in photosphere) the 
accumulated cool material moved already $2-3$~Mm downward, at $t=34$ hours 
(end of right spot formation) temperature difference has recovered to 
pre-emergence values. This effect is limited to the right spot as it is a 
direct consequence of the persistent upflow at the bottom boundary during 
the flux emergence. The upflow inhibits the drainage of low entropy material
and leads to the built up of a top-heavy stratification. The potential energy
stored in that stratification is released when the upflow decays away and
can be used for the amplification of magnetic field in the spot on the right.
 
\subsection{Spot fragmentation}
\begin{figure*}
\includegraphics[width=\textwidth]{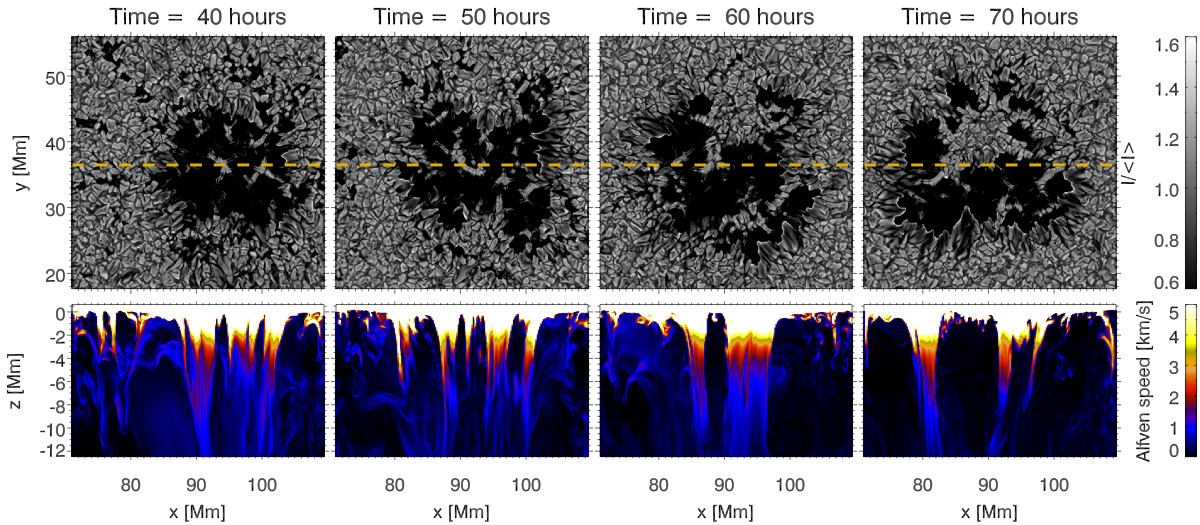}
\caption{Fragmentation of a spot by subsurface convective flows. The time sequence above show the surface brightness of the right spot (upper panels) and vertical cross-section of the associated magnetic structure at and beneath the photosphere (lower panels). Vertical cross-sections are taken at the location of the dashed lines shown in the upper panels. Illustrated in terms of the local Alfv\'en speed, the lower panels show how relatively field-free convective upflows intrude into the spot and eventually carves the spot into fragments. The arrival of such upflows at the photosphere is marked by the initial appearance of a light bridge. As the upflow persists, the light bridge transitions into a granulation pattern more typical of the quiet Sun and plage regions.}
\label{fig:f9}
\end{figure*}
\label{sect:subsurf}
As noted by~\citet{McIntosh:BirthAndEvolutionOfSunspots}, most sunspots (in terms of the presence of an umbra) 
transition to the decay phase almost immediately after formation. Whether this transition is immediate or after a
short stationary phase depends strongly on the quantity considered. In terms of the unsigned flux in the photosphere, we find an immediate transition from formation to decay (see Figure \ref{fig:f2}). In terms of the signed flux, the immediate transition occurs for the left spot but we find a short stationary phase lasting $\sim 30$~hours for the more coherent spot on the right. \citet{Verma:DecayingSunspot} found an immediate decay in terms of active region area for NOAA 11126, while the flux content around the spots showed a stationary phase of about $3$~days.

During the decay phase the shapes of the spots show a systematic deformation and fragmentation leading to breakup. The snapshot at $t=70$~hours (Figure \ref{fig:f1}) shows a more or less ring-like appearance of both spots due
to upflows pressing weakly magnetized plasma into the photosphere. At later
stages ($t=140$~hours) the photospheric magnetic field becomes organized
in larger ($\sim 20$~Mm) cells.

Figure \ref{fig:f7} presents the connection between the photospheric 
vertical magnetic field and the vertical flow velocity in $12$~Mm depth.
White contours in the right panels indicate vertical field exceeding $1.5$~kG 
in the photosphere for better comparison.
 
At $t=70$ hours both spots show clear signatures of deformation and exhibit a
ring-like shape. In both cases the vertical velocity pattern in $12$ Mm depth
shows an upflow cell underneath the spots that sweeps flux into the adjacent
ring-like downflow lanes. 
Stronger field concentrations in the surrounding plage region show  a 
preferred location above downflow lanes, which become more prominent at 
$t=105$ and $140$~hours: the photospheric field distribution approximately 
outlines the vertical velocity structure in $12$~Mm depth. 
For the snapshot at $t=140$~hours we find that about $75\%$ of regions with
more than $2$~kG field strength in the photosphere are above downflow regions at
$12$~Mm depth. For a uniformly random distribution, one would expect this value to be equal to the filling factor of downflows at that depth, which is $39\%$.
Similar to Figure \ref{fig:f5} most smaller scale photospheric flux concentrations are found in downflows of the large scale photospheric velocity field. The most coherent
spot-like flux concentrations are also surrounded by downflows. Since we do not have in our setup penumbrae, these downflows are the consequence of converging
flows in the proximity of the spots as described in detail in \citet{Rempel:SunspotSubsurfaceStructure}.

The connection between the field structure at different depth layers is highlighted in Figure \ref{fig:f8}
where we present the vertical magnetic field for the snapshot at $t=70$ hours at six horizontal cuts in $0$, $2$, $4$, $8$, $12$ and $15$ Mm depth. Also most of the longer living strong field concentrations in the plage region can be traced through most of the domain. In the deepest layers of the domain the convective cell structure is less visible due to influence from the bottom boundary condition.

Figure \ref{fig:f9} presents 4 snapshots during the decay phase of the
right spot that highlight the connection between intrusions of relatively
field free convective flows in the deeper layers and the fragmentation visible
in the photosphere. As the convective intrusion grows in size we find expanding 
patches of granulation appearing in the photosphere that lead ultimately to the
breakup of the spot. Depending of the scale and aspect ratio of the intrusion they
may also lead to the formation of light bridges with a central dark lane as found in the 
simulation of \citet{Cheung:ARFormation}.
This mode of spot fragmentation is similar to the findings in
\citet{Rempel:SunspotSubsurfaceStructure} and not strongly dependent on the initial
state of the simulations. In contrast to our flux emergence simulation,
\citet{Rempel:SunspotSubsurfaceStructure} did start their simulations from strictly
monolithic sunspot models and observed a similar breakup of spots after running
the simulation for more than 24 hours. An overall comparable time scale for the process
is expected since both investigations are based in simulations with the same domain depth.

\subsection{Flow structure in the vicinity of the right spot}
\begin{figure}
  \resizebox{0.95\hsize}{!}{\includegraphics{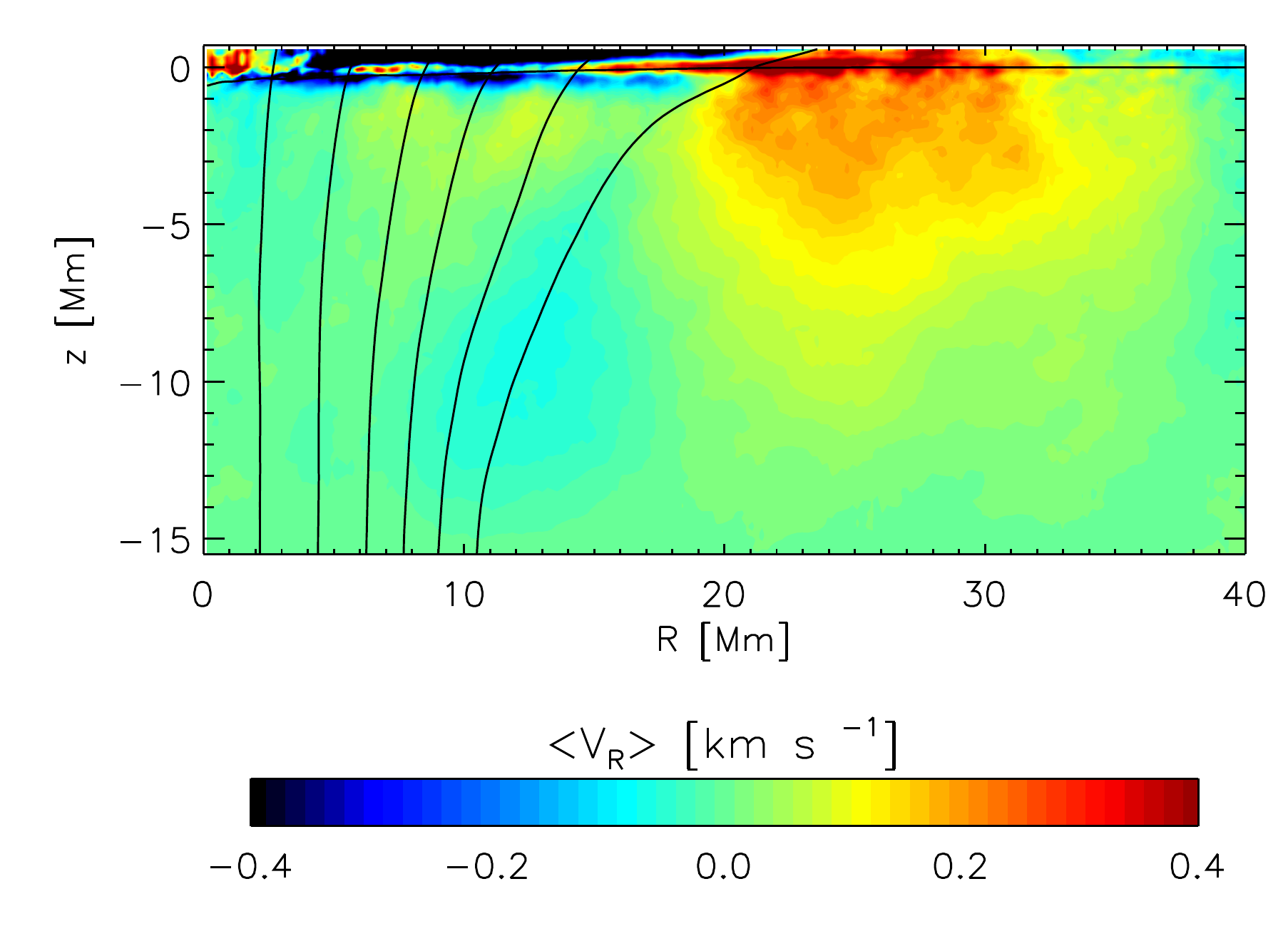}}
  \caption{Azimuthally averaged flows in the radial direction with respect to the approximate center of the right spot. The flows are
    in addition averaged in time from $t=40$ to $t=70$~hours, which corresponds to the nearly stationary phase in the spot evolution
    (Figure \ref{fig:f2}). Color shades indicate the radial velocity amplitude in the range from $-400$ to $400\,\mbox{m s}^{-1}$, with 
    red colors (positive values) indicating outflows away from the spot. We indicate the flux surfaces enclosing 
    $[0.05, 0.2, 0.4, 0.6, 0.8, 1]\cdot 10^{22}$~Mx, the horizontal line indicates the $\tau=1$ level. The spot is surrounded by a 
    region with outflowing material that extends to several Mm of depth and about 2 times the spot radius. The fasted velocities are 
    found in the photosphere where they reach up to $600\,\mbox{m s}^{-1}$.
  }
  \label{fig:f10}
\end{figure}
During the almost stationary phase in terms of signed flux (Figure \ref{fig:f2}) from $t=40$ to $t=70$~hours the right spot
is surrounded by an outflow regions that extends to about $2$ spot radii. We present in Figure \ref{fig:f10} an
azimuthal and temporal average of the radial velocity with respect to the approximate center of the spot over this time period. 
We find at $\tau=1$ an outflow from $R=15$ to about $R=33$~Mm. The peak flow velocity reaches about $600\,\mbox{m s}^{-1}$ in the
photosphere. The amplitude drops rapidly beneath the photosphere, although we can find an outflow down to the bottom of the domain.
Since the spot does not have a penumbra and associated Evershed flow, this flow is clearly of independent origin. While it is possible 
that the outflow is a remnant of the diverging flows that were present during earlier times of the simulation as a consequence of
flux emergence, the overall
flow pattern is very similar to the flows discussed in \citet{Rempel:SunspotSubsurfaceStructure}, despite the very different 
setup (initialization with axisymmetric magnetic field instead of flux emergence). \citet{Rempel:SunspotSubsurfaceStructure} showed that
this flow has two components. While the fast shallow flow in the photosphere is caused by overshooting convection interacting with the inclined 
overlying magnetic canopy, the deeper reaching flow components can be understood as a convective response to the presence of the sunspot. Qualitative
similar flows can be found around cone-shaped obstacles \citet{Rempel:SunspotSubsurfaceStructure} and are a combination of geometric
effects and the modification of the convective heat transport by such obstacles. Within the spot itself we find inflows. They result
from flows directed from granulation towards "naked umbra'' in our setup. Since the spot is not exactly axisymmetric and shows at $t=70$~hours
already a rather complicated structure (see Figure \ref{fig:f9}), the azimuthal average shows this flow to be present within the azimuthally averaged spot. \citet{Rempel:SunspotSubsurfaceStructure} presents a more detailed discussion of this feature and also shows that the inflow disappears in a more coherent
setup with a penumbra. An inflow toward naked umbrae is commonly seen in observations
of pores or naked sunspots \citep{Sobotka:etal:1999,VDominguez:etal:2010,SainzDalda:2012:naked}.  In addition these observations also show divergent flows
(outflows) further away from the pore/naked spot. These flows are typically not referred to as "moat flows'', instead they are described as "outward flows 
originating in the regular mesh of divergence centers around the pore'' \citep{VDominguez:etal:2010}.

In terms of amplitude and radial extent in the photosphere the outflow component we find is consistent with observed moat flows 
\citep[see, e.g.,][]{Brickhouse:Labonte:1988:moat}. The subsurface structure of this flow has been recently studied through
helioseismology by \citet{Featherstone:2011:DeepMoatFlow}. They found in addition to the near surface component with the
strongest amplitude evidence for deeper reaching flows with a secondary peak in about $5$~Mm depth. While we see a significant
depth extent of the outflow, we don't see a secondary peak beneath the photosphere.  

A characteristic feature of the moat region of observed sunspots are moving magnetic features (MMFs) \citep{Harvey:1973:MMF}. We
do not analyze this aspect here in detail, but point the interested reader to the animation of Figure \ref{fig:f1} provided in the
online material, which shows some evidence of patches with both polarities moving away from the right spot after $t=40$~hours. A detailed analysis of the flux transport mechanisms during the growth and decay phases of the spot is given in Section \ref{sect:emf}.

\subsection{Mechanisms for Magnetic Flux Transport During the Formation and Decay Phases of Spots}
\label{sect:emf}
\begin{figure*}
  \centering
  \includegraphics[width=\textwidth]{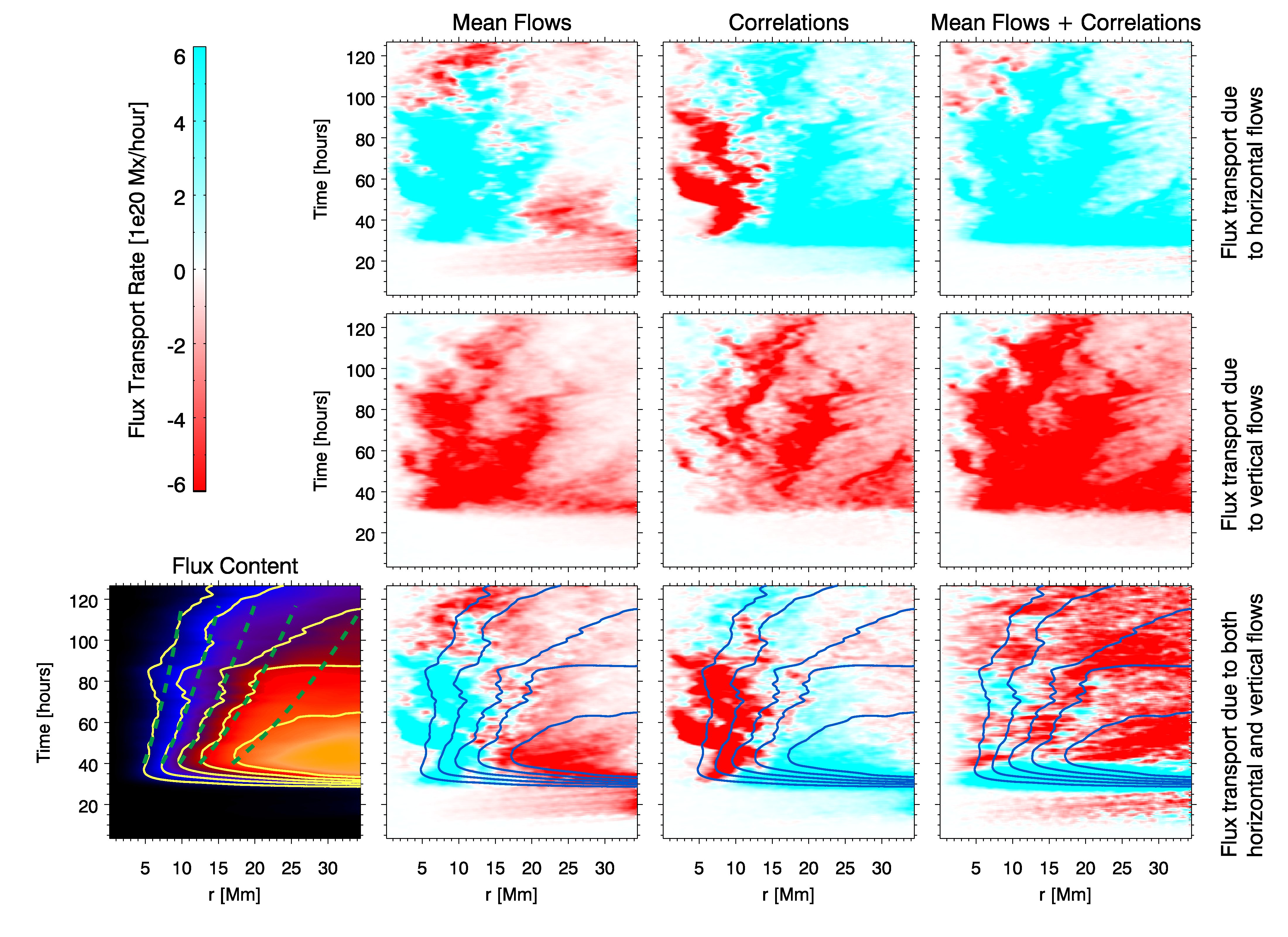}
  \caption{Transport of photospheric ($z=0$) magnetic flux for the right (positive polarity) spot. The bottom left panel shows the net flux content within a circle of radius $r$ from the axis of the right spot. The yellow contours correspond to enclosed net vertical fluxes of $\Phi=2\times 10^{21}$ Mx to $\Phi=10^{22}$ Mx in increments of $\Delta \Phi=2\times 10^{21}$ Mx. Green dashed lines in this panel show the migration of flux surfaces for a self-similar solution for the decay of an ideal spot due to turbulent diffusion. The remaining panels show the flux transport terms, with a positive (negative) value at radial distance $r$ at a certain time $t$ indicating that the flux $\Phi$ enclosed a circle of this radius is increasing (decreasing) in time. The column labeled `Mean Flows' shows the flux transport rate $\dot{\Phi}_{\rm m}$, which is due to azimuthally-averaged  flow $\overline{\mathbf{v}}$ acting on the corresponding mean field $\overline{\mathbf{B}}$. The column labeled `Correlations' shows $\dot{\Phi}_{\rm f}$ due to correlations between fluctuating components $\mathbf{v}'$ and $\mathbf{B}'$. The first and second rows show components of $\dot{\Phi}$ associated with horizontal (i.e. lateral transport) and vertical flows (i.e. emergence and submergence), respectively. The third row shows the sum of both contributions. The plot in the bottom right shows $\dot{\Phi}=\dot{\Phi}_{\rm m } + \dot{\Phi_{\rm f}}$ multiplied by a factor of $4$ to enhance contrast.}
  \label{fig:f11}
\end{figure*}

To examine how magnetic flux grows during the spot formation phase and how it subsequently decays, consider the magnetic field and velocity distribution about the axis of one of the spots in the simulation. Let us decompose the distribution of $\mathbf{B}$ and $\mathbf{v}$ into azimuthally-averaged and fluctuating components (e.g. $\overline{\mathbf{B}}$ and $\mathbf{B}'$, respectively for the magnetic field). For a circle of radius $R$ with boundary $\partial C$ centered on the spot axis (say at a height of $z=0$), it can be shown that the total magnetic flux $\Phi(R)$ crossing $C$ evolves according to~\citep{Cheung:ARFormation}
\begin{eqnarray}
\dot{\Phi}(R) &=& \dot{\Phi}_{\rm m} + \dot{\Phi}_{\rm f}, {\rm~where}\\
\dot{\Phi}_{\rm m}(R) &=& 2\pi R (\overline{v_z}\overline{B_r} - \overline{B_z}\overline{v_r}),\label{eqn:phi_mean}\\
\dot{\Phi}_{\rm f}(R) &=& 2\pi R(\overline{v_z' B_r'}-\overline{B_z' v_r'}).\label{eqn:phi_corr}
\end{eqnarray}

\noindent $\dot{\Phi}_{\rm m}$ indicates magnetic flux transport associated with two effects. The first term in Eq. (\ref{eqn:phi_mean}) is the advection of a radially-directed mean field $\overline{B_r}$ by a mean vertical flow $\overline{v_z}$. The second term is due to the advection of vertical mean field ($\overline{B_z}$) by a mean radially-directed flow ($\overline{v_r}$). The negative sign in front of the second term  indicates that the effect of a positive radial flow is to remove flux from the spot.

In addition to mean field transport by mean flows, correlations between the fluctuating components of $\mathbf{v}$ and $\mathbf{B}$ also play a key role in the flux budget of the simulated spots. Correlations between the fluctuating components leads to an effective electromotive force which gives the flux transport term $\Phi_{\rm f}$ as given by Eq. (\ref{eqn:phi_corr}). Similar to $\Phi_{\rm m}$, this flux transport term includes two effects, namely the vertical advection of radially-directed field ($\overline{v_z' B_r'}$) and the radial advection of vertically-directed field ($-\overline{B_z' v_r'}$). As we previously reported in~\citet{Cheung:ARFormation}, this turbulent flux transport term is associated with the action of granular flows on emerging horizontal fields. Granular flows undulate emerging field lines and expel emerged magnetic flux such that opposite polarities, on average, stream in opposite directions. This effect is key to allowing magnetic flux to migrate from the periphery of the emerging flux region toward the developing spot.

Figure~\ref{fig:f11} shows plots of $\dot{\Phi}$, $\dot{\Phi}_{\rm m}$ and $\Phi_{\rm f}$ for the right (positive polarity) spot for both the formation and decay phases. The flux transport terms are decomposed by their association with vertical and/or horizontal flows. There are some lessons to draw from this figure. To a large extent (especially for $t\ge 40 $ hr, i.e. in the decay phase of the spot), the flux transport terms due to mean ($\dot{\Phi}_{\rm m}$)  and fluctuating components ($\dot{\Phi}_{\rm f}$) are of opposite sign and of roughly equal amplitude. In the mean-field frame work, mean magnetic field evolution results from the net sum between these terms, with the sum generally having much smaller amplitude than the individual mean and fluctuating contributions. In addition also the terms due to horizontal and vertical flows are of comparable amplitude and mostly opposite sign. Imbalances between these terms may depend sensitively on the physical scenario and so one must be careful in drawing general conclusions, in particular when considering individual terms and not the sum of all of them.

With this cautionary note, we point out a few observations from Fig~\ref{fig:f11}. First of all, the contributions due to vertical flows for both $\dot{\Phi}_{\rm m}$ and $\dot{\Phi}_{\rm f}$ are predominantly negative, which means that {\it on average}, both the mean and fluctuating components tend to remove flux from the spot by pumping flux downward. However, the propensity for vertical flows to pump magnetic flux downward does not prevent the spot from accumulating magnetic flux. Inspection of the contributions to $\dot{\Phi}$ due to horizontal flows reveals that flux transport by radially-directed horizontal flows overcompensates to give a net growth in magnetic flux within the spot. At the very early stages of the spot formation process (before $t=28$ hours), diverging horizontal mean flows transport flux away from the center of the flux emergence region, this transport is partially compensated by correlations between $v_r'$ and $B_z'$. In the following spot formation (from $t=28$ to $t=40$ hours), the quantity $\dot{\Phi}_{\rm f}$ is responsible for the inward transport of flux for $r>10-15$~Mm (with the dominant positive contribution from $-\overline{v_r'B_z'}$). Correlations from counter-streaming polarities not only provide an increasing amount of positive polarity flux for the spot, they also expel negative polarity flux to the following polarity spot as well to the boundary of the emerging flux region. Flux transport due to an inward directed mean radial flows becomes a dominant term for the interior regions of the developing protospot ($r<20$ Mm). Some of that is offset by contributions from vertical mean flows so that $\dot{\Phi}_{\rm m}$ is accumulating flux mostly for $r<10-15$~Mm. The converging radial flows are caused downflows that are driven in the center of the protospot as a consequence of an accumulation of cool material, which was prevented from draining during the flux emergence phase (see Figure \ref{fig:f6}). The mean, horizontally converging inflow ($\overline{v_r} < 0$) and the mean downflow ($\overline{v_{z}}<0$) takes the flux previously transported inward by turbulent correlations (i.e. $\dot{\Phi}_{\rm f}$) and concentrates the field to umbral strengths.

In the decay phase of the right (positive polarity) spot ($t>40$~hours), the amplitude of $-\overline{v_r}\overline{B_z}$ and $-\overline{v_r'B_z'}$ (which tends to add flux to the spot) drops and the terms $\overline{v_z}\overline{B_r}$ and $\overline{v_z'B_r'}$ start to dominate. This leads a net removal of flux from the spot, albeit at a slower (unsigned) rate than during the formation phase. A robust trend of this decay phase is that while horizontal flows tend to accumulate flux toward the spot, vertical flows tend to erode it.

The behavior described above may appear to contradict the result of~\citet{Kubo:ARDecay}, who used observations from the Hinode Solar Optical Telescope~\citep[SOT;][]{Kosugi:Hinode,Tsuneta:SOT} to infer flux loss rates of sunspots due to radially directed (relative to the spot axis) outflows. In order to infer the flux loss rate,~\citet{Kubo:ARDecay} used high cadence proxy magnetograms (defined as the ratio of Stokes V and I signals) of a decaying sunspot taken by the narrowband filter imager (NFI) of SOT. Local correlation tracking (LCT) was applied to these high-cadence magnetograms to retrieve horizontal flow maps ($\mathbf{v}_{h,\rm LCT}$) of magnetic patches. The radial component of this was then used to determine the  rate of flux transport by radial flows.  It is important to keep in mind that LCT velocities of line-of-sight magnetic patches do not necessary represent actual plasma velocities that cause temporal changes in magnetograms. As argued by~\citet{DemoulinBerger:EnergyAndHelicityFluxes}, the vertical transport of a magnetic field inclined with respect to the vertical (or line-of-sight) direction would lead LCT to infer an effective horizontal flow. This means that the flux transport rate (denoted $F_v$ in their paper) measured by~\citet{Kubo:ARDecay} may well include contributions from vertical flows acting on horizontal fields ($\overline{v_z}\overline{B_r}$ and $\overline{v_z'B_r'}$).

The simulated spots do not develop an extended penumbra with Evershed flows. It was
shown by \citet{Rempel:Robustness} that penumbrae develop in numerical 
simulations only if a sufficiently inclined field is imposed from the top
boundary in combination with sufficient high resolution (see also Sect. 
\ref{sect:setup:scope}). In the simulations
presented here we do not fulfill both criteria. In order to assess
how the quantities presented in Figure~\ref{fig:f11}
would be affected in the presence of a penumbra we 
computed the equivalent quantities for a high resolution sunspot model
with penumbra. We found no significant differences at the $z=0$ level
considered here, since the Evershed flow has its peak amplitude in the
deep photosphere. In the penumbra $z=0$ shows mostly inverse Evershed flows, 
which are not too different from the flows found in our simulations without 
penumbra.

\subsection{A Simple Model for Spot Decay}

As discussed previously, the loss of magnetic flux from the simulated spots in the decay phase is an complex process involving both mean and turbulent contributions. Furthermore, both vertical and horizontal flows arising from the magnetoconvective system within and near the spots contribute to the net flux transport into and out of the spot. Could a simple  dimensional models of spot decay still be relevant for describing the general behavior of spot decay?

Following~\citet{Meyer:GrowthAndDecayOfSunspots}, consider the photospheric distribution $B_z(r,t)$ of an axisymmetric spot as a function of the radial distance from the axis and time $t$. Assume that at $t=0$, $B_z$ is described by a Gaussian function. If vertical gradients were ignored (without justification) and the field evolves purely under the action of turbulent diffusion with a {\it constant} turbulent diffusivity $\eta_{\rm turb}$, it can be shown that the field would evolve in a self-similar fashion, such that
\begin{eqnarray}
B_z(r,t) &=& \frac{\Phi_0}{\pi \sigma(t)^2 } e^{-r^2/\sigma(t)^2}, {\rm~ where}\label{eqn:self-similar}\\
\sigma(t) & = & \sqrt{\sigma_0^2 + 4\eta_{\rm turb} t}. \label{eqn:ss_sigma}
\end{eqnarray}
\noindent $\sigma_0$ is the width of the Gaussian at $t=0$ and $\Phi_0 = 2\pi\int_0^\infty B_z(r,t=0) rdr$ is the total magnetic flux of the initial spot (which is conserved in time). If the initial profile were a Dirac-delta function such that $\sigma_0=0$ (keeping $\Phi_0$ constant), Eq. (\ref{eqn:self-similar}) would reduce to the solution given by~\citet{Meyer:GrowthAndDecayOfSunspots}. A similar approach for estimating turbulent diffusivities in numerical simulations was also
presented by \citet{Hotta:2012:turbdiff}.

In the bottom left panel of Fig.~\ref{fig:f11}, the green dashed lines show how flux surfaces ($2\times10^{21}$ to $1\times10^{22}$ Mx in increments of $2\times10^{21}$ Mx) would migrate under self-similar evolution. The solution chosen corresponds to $\Phi_0 = 1.1\times10^{22}$ Mx, $\sigma_0 = 11$ Mm, and $\eta_{\rm turb} = 350$ km$^2$ s$^{-1}$. For this solution, $t=40$ hours was chosen as the initial condition of the decaying spot. Between $t=40$ and $t=90$ hours (roughly two turbulent diffusion times), this simple 1D solution provides a somewhat decent match to the average field in the simulated spot. The discrepancy becomes progressively worse at higher flux content. After $t=90$ hours, the numerical solution deviates from the self-similar solution. This maybe attributed to the significant fragmentation of the original spot into smaller pores and to the distortion of its shape (see Fig.~\ref{fig:f8}).

The value of $350$ km$^2$ s$^{-1}$ for $\eta_{\rm turb}$ falls within the range of values of $200-400$ km$^2$ s$^{-1}$ reported by~\citet{Mosher:PhD} in his doctoral thesis examining the decay of ARs in terms of a surface random walk of magnetic field lines.~\citet{Cameron:MixedPolarity} carried out a series of radiative MHD simulations using MURaM code to examined how mixed polarity fields at the surface decay. They reported that the decay of the fields (of an initial uniform strength of $200$ G) was in accordance with an effective turbulent diffusivity of $100-340$  km$^2$ s$^{-1}$. The range in values is a result stemming from the choice of the top boundary condition (in their case, $600$ km above optical depth unity). The use of a potential field boundary condition (as in our case) as opposed to a vertical field boundary condition resulted in an $\eta_{\rm turb}$ near the higher end of their range. Since a potential field boundary condition was used in this the present study, our value of $\eta_{\rm turb}=350$  km$^2$ s$^{-1}$ is not inconsistent with their results. However, we reiterate that our self-similar solution for the spot decay using this value of $\eta_{\rm turb}$ only applies for the first two days of the decay phase ($t=40$ to $t=90$ hours). In latter stages, the fragmentation of the spots (driven by deep subsurface flows) accelerates the decay of the spot. Since the simulations of~\citet{Cameron:MixedPolarity} used a computational domain that only extended to $800$ km below optical depth unity and the horizontal extent of their domain was limited to $6\times6$ Mm$^2$, their simulations do not capture the effects of larger-scale flow patterns in deeper layers.

The assumption of a constant turbulent diffusivity $\eta_{\rm turb}$ is a simplification that keeps the diffusion equation linear to facilitate analytical solutions. This implicitly assumes that the erosive effects of turbulent diffusion do not depend on field strength. Since it is well known that strong magnetic fields (e.g. in sunspots) inhibit convective motions~\citep[see, e.g.][]{Schuessler:UmbralConvection}, this assumption is not valid. Without performing full 3D MHD simulations (which is what we do here), a appropriate treatment would be that taken by \citet{PetrovayMorenoInsertis:TurbulentErosion}, who took into account the strong-field quenching of turbulent diffusion by assuming an ad hoc (but physically motivated) functional dependence of $\eta_{\rm turb}$ on field strength $B$. In fact, observations of sunspot decay typically require lower turbulent diffusivities~\citep[e.g. $200$ km$^2$ s$^{-1}$, ][]{MartinezPillet:SunspotDecay} than those of global flux transport models, which follow the evolution of magnetic flux over much longer time scales (months, years and solar cycles) typically require a larger turbulent diffusivity of $\eta_{\rm turb} \sim 600$ km$^2$ s$^{-1}$~\citep{Sheeley:LivingReviews}. These higher values of the turbulent diffusivity are usually associated with supergranulation rather than granulation. Since there are no (clear) signatures of supergranulation in our model, the influence of supergranular cells on the AR decay process is not addressed in the present work.

\begin{figure*}
\includegraphics[width=\textwidth]{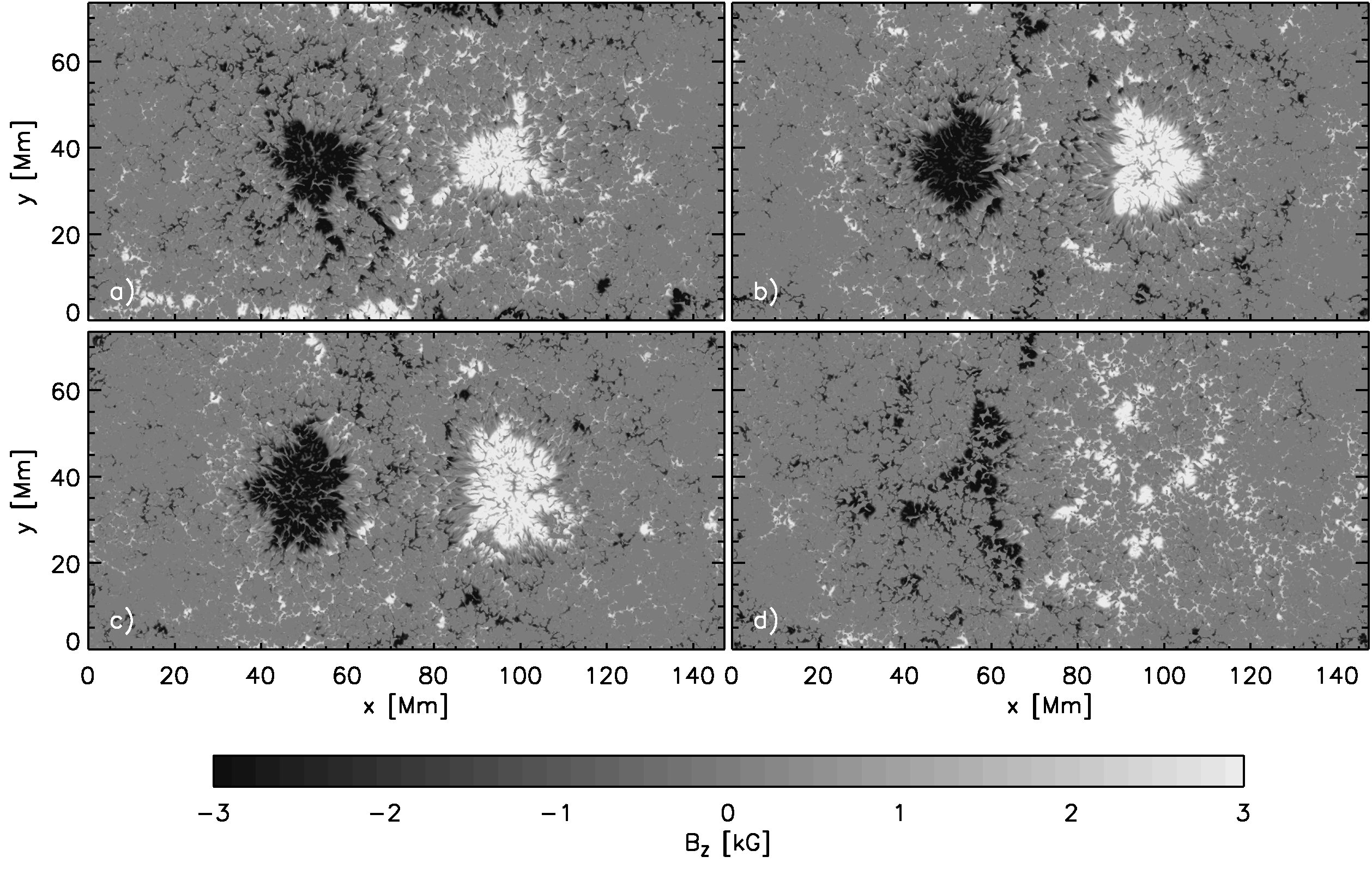}
\caption{Comparison of results from 4 different flux emergence simulations. The top panels compare two simulations with different initial field strength, but the same overall flux:  a) 10.6 kG; b) 21.2 kG. In both cases the bottom boundary condition after the flux emergence is an open boundary as described above. In panels c) and d) the initial field strength is
21.2 kG, but we used different bottom boundary conditions. In panel c) we allow for horizontal flows, but no vertical flows in regions with more than 2.5 kG field strength, in panel d) we continued the inflow in regions with more than 2.5 kG field
strength. Except for panel d), where we did not observe spot formation, results are comparable
}
\label{fig:f12}
\end{figure*}

For the self-similar solution given by Eqs. (\ref{eqn:self-similar}) and (\ref{eqn:ss_sigma}), the flux contained with a radius $r$, and the flux change rate are, respectively, given by
\begin{eqnarray}
\Phi(r,t) &=& \Phi_0 \left[1-e^{-r^2/\sigma(t)^2}\right],\\
\dot{\Phi}(r,t)=\frac{\partial \Phi}{\partial t}& = & -\Phi_0\frac{4\eta_{\rm turb}r^2}{\sigma(t)^4}e^{-r^2/\sigma(t)^2}.\label{eqn:ss_flux_loss}
\end{eqnarray}
For the present case, Eq.~(\ref{eqn:ss_flux_loss}) yields flux {\it loss} rates as high as $4.7\times10^{16}$ Mx s$^{-1}$ at the beginning of the decay phase. Depending on $r$ and $t$, the flux loss rate can be many times smaller. This result is consistent with the sunspot flux loss rates reported by~\citet{Kubo:MMFs} for an actual sunspot. Their reported values are in the range $2.3-5.0\times 10^{21}$ Mx day$^{-1}$ (corresponding to $2.6 - 5.8\times 10^{16}$ Mx s$^{-1}$). It should be noted that the sunspot they studied initially had twice the flux of each of our simulated spot. A study of sunspot decay by~\citet{MartinezPillet:SunspotDecay} found that the rate of flux decay from sunspots is of order $10^{20}$ Mx day$^{-1}$, which is one order of magnitude smaller than the flux loss rate for our simulated spot and the flux loss rates reported by~\citet{Kubo:MMFs}.~\citet{Kubo:MMFs} speculated that the discrepancy arises from the difference in flux contents of the spots (with the spots studied by~\citet{MartinezPillet:SunspotDecay} having a flux content 2-3 times smaller than the one they observed). Regardless of the possible origin of the discrepancy in measured flux loss rates in observed sunspots, we caution that the flux loss rate in our simulation may be artificially high due to the limited depth captured by the computational domain.

The self-similar solution of the decaying spot can also be used to study the apparent rate at which the flux spreads over the solar surface. As should be apparent in Eq. (\ref{eqn:self-similar}), $\sigma(t)$ represents the characteristic width of the Gaussian profile at any time $t>0$. Most ($86\%$) of the original flux of the spot is contained within $r<\sigma(t)$. Taking this as an effective boundary for the flux content of the diffusing structure,
\begin{equation}
\frac{d\sigma}{dt} = v_{\rm boundary}=\frac{2\eta_{\rm turb}}{\sqrt{\sigma_0^2+4\eta_{\rm turb}t}}
\end{equation} 
\noindent may be interpreted as the effective propagation speed of the flux boundary across the solar surface. For the simulated spot, $v_{\rm boundary} = 64$ m s$^{-1}$ at the beginning of the decay phase ($t=40$ hours in the simulation). After $50$ hours of decay, $v_{\rm boundary}$ diminishes to $36$ m s$^{-1}$. While these numbers should only be considered estimates on the effective propagation speed of the diffusing magnetic flux, they are nevertheless interesting for the question of the long term stability of active regions. Helioseismic measurements of near-surface flows indicate the existence of persistent inflows into AR centers~\citep{Haber:SubsurfaceInFlows}. The amplitude of these inflows is reported to be $20-30$ m s$^{-1}$~\citep{Hindman:SubsurfaceCirculationWithinARs}. The existence of these inflows could mean that when $v_{\rm boundary}$ becomes sufficiently small, the further spreading of an AR is at least partially suppressed by inflows.

\subsection{Robustness of the Spot Formation Mechanism}
\label{sect:controlexp}

To examine the robustness of the various physical processes described above associated with spot formation, we carried out some additional simulations with varying field strengths and boundary conditions. In Figure \ref{fig:f12} we present
magnetograms at the $\tau=1$ surface for 4 different control experiments. The snapshots shown are taken about 8 hours after
the spot formation started in the respective simulation. In contrast to the simulation presented above all these experiments did not have a field aligned flow imposed, leading to spot pairs with no significant asymmetry. Panel a) shows a
control experiment with initially 10.6 kG field strength, which is closest to the simulation presented here in detail. In panel b)
we increased the initial field strength to 21.2 kG, leading to a spot pair with almost the same photospheric appearance.
The increased initial field strength does not influence the field strength of the spots, but leads to moderately more
coherent spots and less flux in the surrounding plage region. In particular the accumulation of flux near the domain 
boundaries is less pronounced. Since the kept the total flux unchanged, doubling the initial field strength cuts the volume
of the torus in half and therefore reduces the overall amount of mass that is forced to overturn during the flux emergence
process. Panel c) presents a simulation similar to panel b), but we transitioned to a different bottom boundary condition after the flux emergence. While we allow in panel b) for both vertical and horizontal flows (symmetric boundary condition on all three mass flux components), we do not allow for vertical flows in regions with more than 2.5 kG in panel c), i.e. the vertical mass flux is antisymmetric across the boundary in regions where $\vert B\vert$ exceeds 2.5 kG, but symmetric everywhere else. The consequence of this setup is that downflows that form during the
spot formation process have to spread out at the bottom boundary and in return destroy the coherence of the magnetic footpoints. The results presented in panels b) and c) do not show significant differences, implying that the coherence of spots in the photosphere is not related to the coherence of field at the footpoints. The most dramatic difference happens when we do not transition after the flux emergence to a boundary without inflows at the footpoints. In panel d) we present the result from a simulation in which an inflow of 500 m/s was kept
in regions at the bottom boundary condition with more than 2.5 kG field strength. In this case we did not observe the formation of a pair of spots. Instead the flux emergence led to a mixed polarity plage region with intermittent pores with the size of granules. While the formation of these small pores may possibly be attributed to properties of near-surface stratified magnetoconvection~\citet{Kitiashvili:Spontaneous, Kemel:SpontaneousFormation}, the combined results from the various simulations here suggests that subsurface properties (such as flows) are critical to spot formation.  

Overall these control experiments illustrate that the most crucial aspect of our setup is the transition from emergence to an open boundary condition that does not impose persistent upflows. Results are not too sensitive on how fast this transition happens as long as the transition time is short compared to the typical convective time scale of the domain (about one day based on the average up- and downflow velocities). We varied the transition time scale between 20 minutes and 6
hours and obtained comparable results. With a longer transition time scale the formation of spots is delayed. 

\subsection{Implications from Absence of Penumbra and Twist}
\label{sect:penumbra_twist}
The simulations presented here do not include penumbrae. Penumbra formation and decay throughout the course of the simulation can be included in
principle through a combination of higher resolution and different top boundary conditions,
however, such a setup would make the simulation at least one order of magnitude more expensive and we decided to exclude those aspects for now. This
raises the question of how our results are potentially influenced by that choice. While most processes related to the formation of spots
happen before a penumbra would form in the photosphere and are therefore likely not affected, this is not necessarily true during the decay 
phase. We find in the presented simulation a rather rapid decay driven by subsurface flows fragmenting the magnetic field on a time scale of
about a day. While similar processes are at work in observed sunspots, the time scale in our simulation is by factors of a few too short, essentially not allowing for spots with lifetimes beyond a few days. While the limited domain depth and the resulting lack of convective time scales 
beyond a few days is certainly a key factor leading to overall short life times, the absence of a penumbra could be another factor.  
There are two ways this could potentially happen: 1. A penumbra could stabilize the near surface layers against decay even if the subsurface layers
strongly fragment. 2. A penumbra could delay the fragmentation of the subsurface layers. Addressing the potential influence of a penumbra on sunspot
decay requires further investigation, which is beyond the scope of this paper, but work in progress. 

To examine the role of twist in the spot formation process, we conducted a few control experiments with lower resolution, which show that the addition of twist
increases the amount of mixed polarity flux emerging in the photosphere and also the amount of flux found later in the spots. In addition, 
twist causes the resulting spots to rotate \citep[see, e.g.,][]{Cheung:ARFormation}, which leads to spots with a higher degree of axisymmetry.
However, the simulation results lead us to conclude that, provided there is sufficient flux that has emerged at the photosphere, twist is not a determining factor in the spot formation process. 

This conclusion should be distinguished from lessons from previous work on the role of twist in maintaining flux tube coherence during their rise in the deep convection zone\citep[see, e.g.,][]{Fan:LivingReview2009}. In the near-surface layers of the convection zone the underlying dynamics differ substantially from the deeper layers due to the diminishing pressure scale heights near the surface. An active region scale flux bundle moves at most a distance comparable
to its own diameter while expanding by a factor of about $10^4$. The strong expansion is mandated by the density stratification and cannot be suppressed by any reasonable amount of twist. Flux emergence into the photosphere and above is only possible if the magnetic field decouples on average from the mass, which is only efficient and fast if the magnetic field becomes temporarily organized on very small scales. As discussed in Section \ref{sect:emf}, the granular convective flows near the surface allows this to happen. This overcomes the need for twist to aid the transport of flux into the solar atmosphere which is found in models that do not include convective flows~\citep[e.g.][]{Toriumi:TwoStepEmergence}.

\section{Discussion}
\label{sect:discussion}
We presented a series of flux emergence simulations similar to \citet{Cheung:ARFormation}.
The main differences are a larger domain with about 3 times more overall flux and an about
$10$ larger overall density contrast. In addition we considered here also a setup with a field 
aligned flow directed in the negative $x$-direction. Our simulations
lead to the formation of a pair of spots with about $10^{22}$~Mx flux, which is
about $50-60\%$ of the flux we moved into the domain across the bottom 
boundary. Interestingly this ratio is not very different from 
\citet{Cheung:ARFormation} despite the fact that we a) considered an untwisted 
flux tube, b) had $10$ times more density stratification in our domain, and
c) started with an initially 2 times weaker magnetic field. Overall this 
points to some robustness of the processes we find during flux emergence
and sunspot formation.

The addition of a field aligned flow, which is expected to arise as a consequence of angular momentum conservation,
leads to two distinct photospheric signatures: 1. We find a significant asymmetry between the two spots. The spot that
would correspond on the Sun to the leading spot is more coherent and axisymmetric. 2. We find systematic differences 
in the timing of the formation of both spots. The spot corresponding to the leading spot on the sun forms overall
faster after an initial delay. While both spots reach about half of their peak flux at the same time, the formation 
of the left spot (corresponding to the trailing spot) starts earlier and continues past the formation of the right spot.
This appears to be opposite to the observed behavior of the majority of ARs, although \citet{McIntosh:BirthAndEvolutionOfSunspots} 
reports on 2 cases (out of 15)
in which the trailing spot formation preceded the leading spot formation by $7$ and $26$ hours (we find about $10$ hours
difference). In a recent observational study of the birth of two ARs with SDO/HMI data, the following spot seems to form earlier in one case (AR 11105) but later in the other~\citep[AR 11211, see Figure in][]{Centeno:NakedAR}. While our simulation may not represent the "typical" case, it does not strictly contradict observations in this aspect either. Comprehensive statistical studies
surveying spot formation times and other observables in leading and following polarities will be crucial to settle the question of what is a typical AR and what are the expected deviations from the typical AR. We note that our setup was chosen to highlight the influence of a field aligned flow in separation, rising flux tube simulations typically lead to additional asymmetries between both polarities that could potentially alter that outcome. We also note that the horizontal extent of the simulation
domain in conjunction with periodic boundaries prevents the spread of magnetic field in the photosphere and potentially 
influences the timing of the left spot formation.

In our setup the total mass enclosed by the semi-torus we emerge across the bottom boundary corresponds to about $50\%$ of the total mass in the domain. 
Consequently continuing flux emergence and spot formation is only possible if mass and magnetic field decouple on average, allowing for the mass to leave
the domain through the open bottom boundary while more than $50\%$ of the emerging flux reaches the top boundary of the domain (the total density contrast
in the mean stratification is more than $10^6$). The decoupling of mass and magnetic field happens mostly through two processes. In the early stages of
flux emergence persistent inflows at the bottom boundary condition lead to the formation of a subsurface field configuration with strong alignment between field
and flow. This is mostly the case for the right spot with the stronger inflow at the footpoint. Due to flux freezing the overturning mass flux initially holds down 
most of the magnetic field and prevents emergence of flux into the photosphere (less than $0.1\%$~of the mass flux present at the bottom boundary reaches the photosphere
due to strong stratification). Since most flows are field aligned, the strong horizontal 
divergence of the velocity field does not weaken the field and leads to the built up of a several kG strong horizontal magnetic layer in the middle of the domain.
At later stages the horizontal field rises toward the surface and emerges through the photosphere in form of granulation scale $\Omega$-loops that allow for an
effective decoupling of mass and magnetic field. Most of the magnetic flux emerges through the photosphere in form of mixed polarity field, spots form after a separation
of polarities.

For both spots the field re-amplification in the photosphere is related to 
strong downflows, although the origin of these downflows differs. The left
spot forms above the downflow region at the periphery of the flux emergence 
region. This downflow region is a consequence of mass conservation, i.e. the
mass which is pushed into the domain through our emergence boundary condition
has to leave the domain as we keep the average gas pressure at the bottom 
boundary unchanged. Since this downflow region shows an orientation in the
y-direction, the initial shape of the left spot is also more sheet-like
than axisymmetric. In the case of the right spot the longer lasting upflow 
prevents the drainage of low entropy material continuously forming in the
photosphere. As a consequence we find and enhancement of the average 
upflow/downflow temperature contrast in about $8-10$~Mm depth beneath the 
photosphere. This effect is strongest in the center, where horizontally
diverging flows are weakest. After the inflow decays away (as a consequence 
of switching the
bottom boundary condition back to an open boundary), the resulting top heavy
stratification leads to strong downflows, while the surrounding
system of strong horizontal field is moving upward. Flux emerges in the
photosphere in form of $\Omega$-loops while the positive polarity field is amplified in
downflows preferentially located above the foot-point of the right spot. The almost
circular shape of the right spot is a reflection of the circular shape
of preceding subsurface flow field with strong horizontal divergence.

In terms of the total unsigned flux the simulated active region transitions directly
from formation to decay phase. For the right spot we find a short $\sim 30$~hours stationary phase
in terms of signed flux between formation and decay.
This behavior is qualitatively similar to observational reports studying
the formation and decay of ARs~\citep[e.g.][]{McIntosh:BirthAndEvolutionOfSunspots,Verma:DecayingSunspot},
but we find substantial differences in the associated time scales.
The decay is mostly driven by the largest scale convective motions present
in our domain, i.e. the convection pattern near the bottom boundary. This
sets both the decay time scale as well as surface magnetic field pattern 
during the decay phase. The time scale of decay is of the order of a few days,
which is also consistent with the single spot simulations of 
\citet{Rempel:SunspotSubsurfaceStructure}, but significantly shorter than the typical life time
of sunspots.
This reinforces their conclusion that realistic spot life times
require simulations in deeper domains, likely around $30-50$~Mm depth.
The photospheric magnetic field pattern during the decay phase approximately 
reflects the convective cell structure near the bottom boundary. In our setup it happened that
new upflow cells formed at the bottom boundary beneath both spots leading to a ring-like appearance
during the decay phase. This is a detail of the simulations that depends strongly on the location of
the bottom boundary and would likely not happen in a substantially deeper domain.

In the initial stages of spot decay, the spreading of vertical flux over the photospheric surface was shown to be reasonably well-described in terms of a turbulent diffusion process. A self-similar solution to the diffusion equation for a Gaussian profile~\citep[][]{Meyer:GrowthAndDecayOfSunspots,Mosher:PhD}with constant turbulent diffusivity was found to approximate the lateral migration of flux surfaces over the course of two days. During this time, the associated flux loss rates (from both the numerical simulation and analytical model) are in accordance with measured flux loss rates from observed sunspots as reported by~\citet{Kubo:MMFs}. Thereafter, fragmentation of the spot led to increased rates of magnetic flux loss.

During formation and decay phase flux transport in the photosphere cannot be easily represented by simple expressions. We find that in
general contributions from mean and fluctuating parts partially offset each other, similarly the contributions from radial and vertical expressions tend to have
opposing signs. In particular the expression $\overline{-v_rB_z}$ alone is insufficient to account for the flux transport rate and the complementary component $\overline{v_zB_r}$ must also be taken into account.

The patterns of spot formation and decay we found in this investigation are 
a consequence of the flux emergence boundary condition we implemented.
This raises the question which aspects can be considered robust and which 
aspects are very specific to this setup. Nevertheless, a comparison to 
\citet{Cheung:ARFormation} already shows that neither twist, overall flux,
field strength nor density contrast (domain depth) have a dramatic influence
on the overall evolution. We have conducted additional control experiments
using the same domain size and overall flux. We did not find a significant
difference between simulations starting with $10$ and $20$~kG initial
field strength. We also explored different bottom boundary conditions
after the initial flux emergence. There is no significant difference between
keeping the magnetic field fixed or allowing the foot-points to erode. The
one aspect which is crucial is a decay of the initial upflow. If the upflow
is continued at the bottom boundary no spot formation is observed. Since flux
emergence is (observationally) a event with limited duration, using a boundary
condition that switches after about $20$ hours of emergence to an open boundary as 
we did seems very reasonable. Identifying the underlying 
physical process is beyond the scope of our numerical setup, but will be addressed
in the future as larger and deeper domains become available.

Recently \citet{SteinNordlund:ARFormation} demonstrated the formation of a small active region in a setup which did not impose initially a coherent flux bundle emerging though the bottom boundary (located at $z=-20$ Mm). Instead they imposed horizontal flux in inflow regions and the formation of a bipolar group occurred as consequence of convective scale selection within the simulation domain. On the one hand, this indicates a substantial degree of robustness with rather weak dependence on the initial field configuration. In light of this, perhaps the photospheric field structure may not tell us much about subsurface field properties. On the other hand, the choice of imposing a uniform horizontal field (with constant strength and orientation over the course of two days) in upflows crossing the bottom boundary implicitly assumes the existence of a large-scale, coherent magnetic structure at depths below $20$ Mm.

Recently \citet{Brandenburg:NEMPISpots} proposed a mechanism for the near
surface field amplification of pre-existing flux based on the so called negative
magnetic pressure instability (NEMPI) that leads to super-equipartition field strengths
regardless of subsurface conditions. While it is difficult to quantify if this mechanism also
contributes to amplification in our simulations, the asymmetry of observed spots on the sun and
the empirical fact the spot formation happens during or just after flux emergence suggests
that NEMPI is not the sole mechanism for spot formation.

A more detailed comparison with flow observations during an active region
formation phase is needed to put better constraints on the subsurface 
magnetic field and flow structure responsible for flux emergence. The 
numerical model presented here is a prototype simulation intended to highlight how 
a combination of upflows and field-aligned flows can influence the flux 
emergence and subsequent sunspot formation process. A detailed study comparing 
flows in the simulation presented here (and other comparable simulations) 
with flows inferred through helioseismology is 
work in progress.

\acknowledgements
The National Center for Atmospheric Research is sponsored by the National 
Science Foundation. This work used the Extreme Science and Engineering 
Discovery Environment (XSEDE), which is supported by National Science 
Foundation grant number OCI-1053575. Computing resources were provided by 
the National Institute for Computational Sciences (NICS) under grant 
TG-AST100005. We thank the 
staff at the supercomputing center for their technical support. MCMC acknowledges computing support from NASA's Advanced Supercomputing Division, which operates the Pleiades cluster at Ames Research Center. MCMC acknowledges support by NASA contract NNM07AA01C to LMSAL and support by the Lockheed Martin SDO/HMI sub-contract 25284100-26967 from Stanford University (through Stanford University prime contract NAS5-02139).
MCMC thanks the participants of the ISSI team for Flux Emergence and attendees of the 2013 Flux Emergence Workshop for stimulating discussions.

\bibliographystyle{natbib/apj}
\bibliography{natbib/references}

\end{document}